\documentclass[11pt]{article}
\usepackage{jcapmod}

\usepackage{booktabs}
\usepackage[english]{babel}
\usepackage{amsmath,amssymb,amsbsy,amstext, amsthm, simplewick}
\usepackage{graphicx}
\usepackage{amsfonts}
\usepackage{amssymb}
\usepackage{upgreek}
\usepackage{simplewick}
 \usepackage{exscale,relsize}

\usepackage[margin=1cm,labelfont={sf,bf,scriptsize},textfont={sf,scriptsize}]{caption}

\makeatletter
\newlength{\apb@width}
\newcommand{\autoparbox}[2][c]{\settowidth{\apb@width}{#2}\parbox[#1]{\apb@width}{#2}}
\newcommand{\includegraphicsbox}[2][]{\autoparbox{\includegraphics[#1]{#2}}}
\makeatother

\numberwithin{equation}{section}

\def\beq{\begin{equation}}
\def\eeq{\end{equation}}

\def\bea{\begin{eqnarray}}
\def\eea{\end{eqnarray}}

\def\d{{\rm d}}

\def\beq{\begin{equation}}
\def\eeq{\end{equation}}
\def\bea{\begin{eqnarray}}
\def\eea{\end{eqnarray}}

\def\d{{\rm d}}

\def\Mp{M_{\rm pl}}
\def\d{{\rm d}}

\def\H{{\cal H}}

\def\0{{\vec{0}}}
\def\k{{\vec{k}}}
\def\q{{\vec{q}}}

\def\x{{\vec{x}}}
\def\y{{\vec{y}}}
\def\z{{\vec{z}}}

\def\p{{\vec{p}}}

\DeclareRobustCommand{\SkipTocEntry}[4]{}

\def\fnl{f_{\mathsmaller{\rm NL}}}

\def\tnl{\tau_{\mathsmaller{ \rm NL}}}

\newcommand{\ket}[1]{| #1 \rangle}
\newcommand{\bra}[1]{\langle #1 |}

\def\Mp{M_{\rm pl}}
\def\d{{\rm d}}

\def\tl{\theta_{\mathsmaller{L}}}
\def\sl{\sigma_{{\mathsmaller{L}}}}
\def\ts{\theta_{\mathsmaller{S}}}
\def\ss{\sigma_{{\mathsmaller{S}}}}
\def\L{\mathsmaller{L}}
\def\SS{\mathsmaller{S}}

\newcommand{\vev}[1]{\langle #1 \rangle}

\def\H{{\rm H}}

\DeclareSymbolFont{extraup}{U}{zavm}{m}{n}
\DeclareMathSymbol{\varheart}{\mathalpha}{extraup}{86}
\DeclareMathSymbol{\vardiamond}{\mathalpha}{extraup}{87}

\begin{document}

\begin{titlepage}

\setcounter{page}{1} \baselineskip=15.5pt \thispagestyle{empty}

\bigskip\

\vspace{2cm}
\begin{center}
{\fontsize{19}{36}\selectfont  \sc On Soft Limits of\\ \vskip 10pt Inflationary Correlation Functions}
\end{center}

\vspace{0.6cm}

\begin{center}
{\fontsize{13}{30}\selectfont   Valentin Assassi$^{\spadesuit}$, Daniel Baumann$^{\spadesuit}$, and Daniel Green$^\clubsuit$}
\end{center}


\begin{center}
\vskip 8pt
\textsl{$^\spadesuit$ D.A.M.T.P., Cambridge University, Cambridge, CB3 0WA, UK}

\vskip 7pt
\textsl{$^\clubsuit$ School of Natural Sciences,
 Institute for Advanced Study,
Princeton, NJ 08540, USA}

\end{center}

\vspace{1.2cm}
\hrule \vspace{0.3cm}
{ \noindent \textbf{Abstract} \\[0.2cm]
\noindent 
Soft limits of inflationary correlation functions are both observationally relevant and theoretically robust.
Various theorems can be proven about them that are insensitive to detailed model-building assumptions. In this paper, we re-derive several of these theorems in a universal way. Our method makes manifest why soft limits are such an interesting probe of the spectrum of additional light fields during inflation. We illustrate these abstract results with a detailed case study of the soft limits of quasi-single-field inflation.}

 \vspace{0.3cm}
 \hrule

\vspace{0.6cm}
\end{titlepage}

\tableofcontents

\newpage
\section{Introduction}

Soft limits of primordial correlation functions are of special phenomenological relevance, since many observational probes of non-Gaussianity---such as scale-dependent halo bias~\cite{Dalal:2007cu} and CMB $\mu$-distortions~\cite{Pajer:2012vz}---involve a separation of scales.
At the same time, soft limits are theoretically clean. 
Focusing on soft limits allows one to prove a number of interesting theorems that are insensitive to detailed model-building assumptions, but are sensitive to the number of light degrees of freedom during inflation. Moreover, as we will explain in this paper, the momentum scaling in the soft limits is sensitive to details of the spectrum of particles during inflation.

\vskip 4pt
There are two classes of soft limits:
\begin{enumerate}
\item The `squeezed limit' of an $N$-point function refers to taking one external momentum to be smaller than all others~(see Fig.~\ref{fig:soft1}).
\begin{figure}[h!]
   \centering
       \includegraphics[scale =0.33]{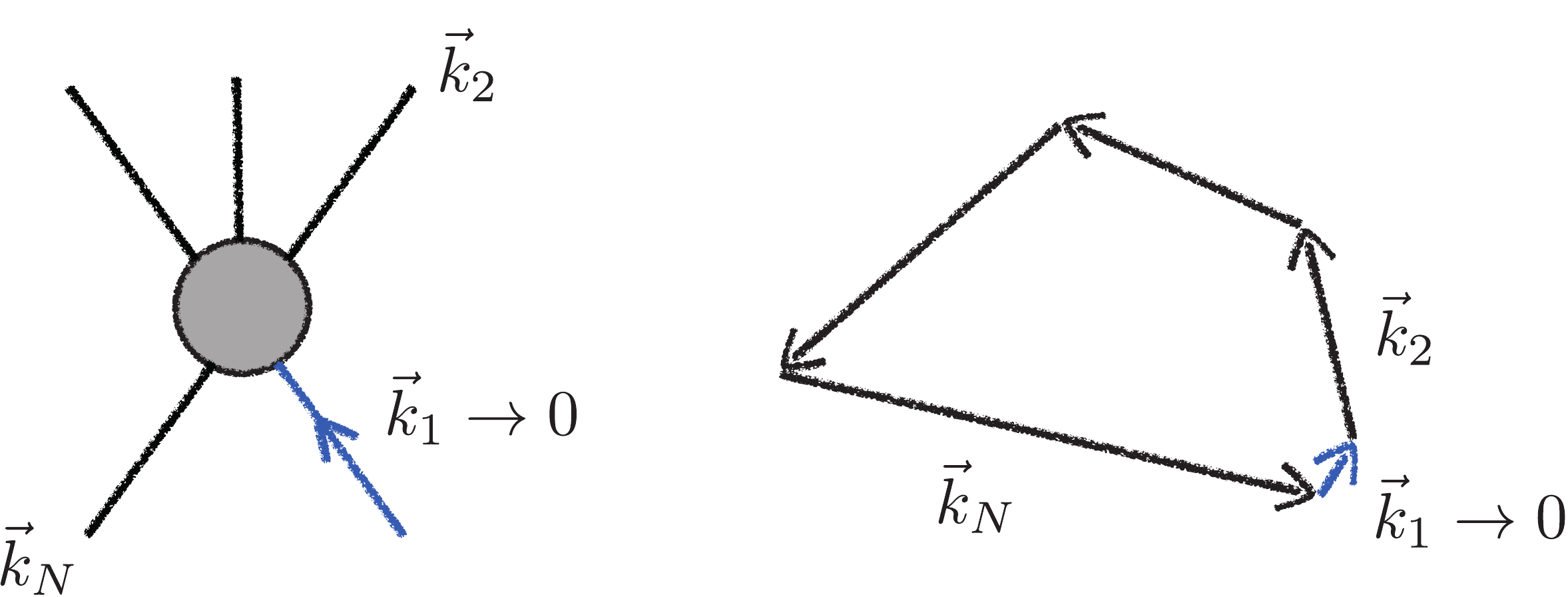}
   \caption{Soft external momentum or the `squeezed limit'.}
  \label{fig:soft1}
\end{figure}
Maldacena's consistency relation for single-field inflation~\cite{Maldacena:2002vr} (see also~\cite{Creminelli:2004yq, Cheung:2007sv,Creminelli:2011rh}) relates the squeezed limit of the three-point function of curvature perturbations $\zeta$ to the variation of the two-point function under dilation. Similar results have subsequently been derived for $N$-point functions \cite{Huang:2006eha, Li:2008gg,Leblond:2010yq, Seery:2008ax}.  These results are explained by realizing that the curvature perturbation $\zeta$ non-linearly realizes a dilation symmetry in any FRW background~\cite{Hinterbichler:2012nm} (see also~\cite{Creminelli:2012ed,Hinterbichler, Goldberger}).
Any violation of the consistency relations would rule out {\it all}\, single-field models and point to additional degrees of freedom during inflation.

\item The `collapsed limit' of an $N$-point function takes one internal momentum (i.e.~the sum of $M$ of the $N$ external momenta) to be smaller than all external momenta (see Fig.~\ref{fig:soft2}).
\begin{figure}[h!]
   \centering
       \includegraphics[scale =0.33]{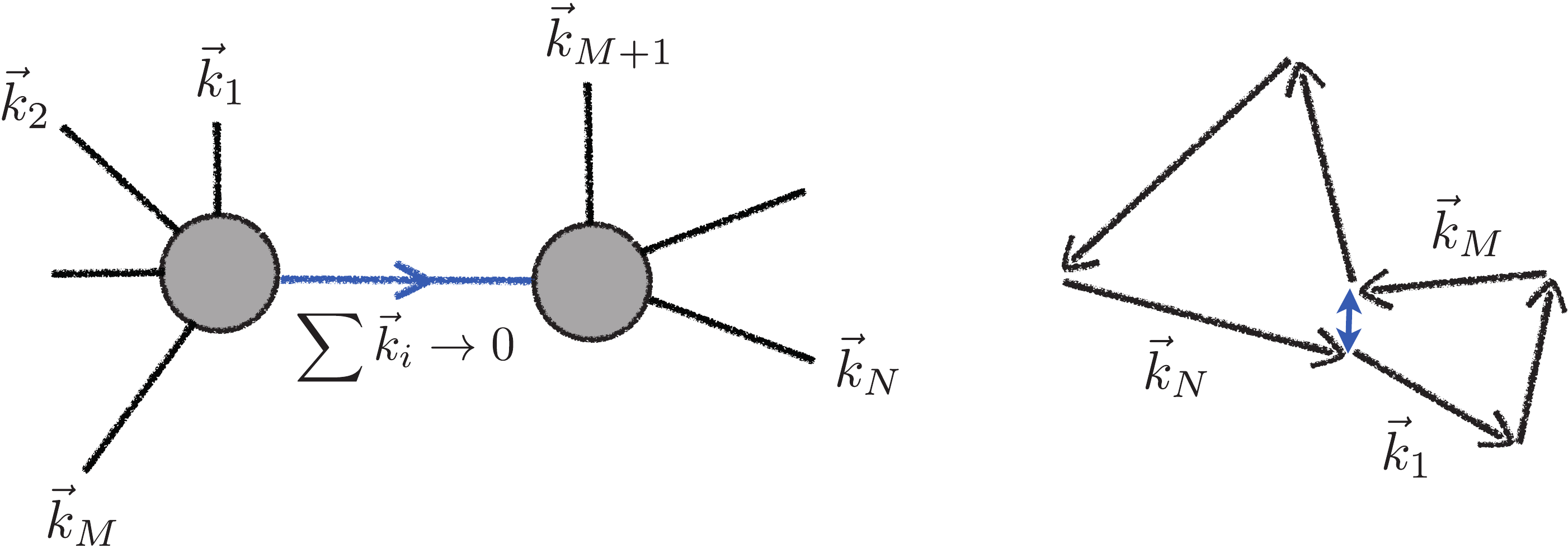}
   \caption{Soft internal momentum or the `collapsed limit'.}
  \label{fig:soft2}
\end{figure}
The Suyama-Yamaguchi inequality~\cite{Suyama:2007bg} (see also \cite{Sugiyama:2011jt,Smith:2011if, Sugiyama:2012tr}) puts a bound on the collapsed limit of the four-point function in terms of the squeezed limit of the three-point function. 
At tree level, the inequality is saturated if a `single source' creates the primordial curvature perturbation and its non-Gaussianity~\cite{Byrnes:2006vq}. Collapsed limits are therefore probes of `multiple source' (or `hidden sector') non-Gaussianities. An interesting observational signature of this effect is stochasticity in the scale-dependent halo bias~\cite{stochastic} (see \cite{Jeong:2012df} for another possible signature).
\end{enumerate}

A boost in the collapsed limits arises naturally in models of quasi-single field inflation (QSFI)~\cite{Chen:2009zp}. 
In these models a light inflaton field mixes with an additional scalar field (the `isocurvaton') with mass of order the Hubble scale $H$.\footnote{Such a spectrum of particle masses arises generically in supersymmetric theories of inflation~\cite{Baumann:2011nk}.} This mixing allows large non-Gaussianities in the isocurvaton sector to be communicated to the observable sector.  The time dependence of the massive field on superhorizon scales leads to a characteristic scaling behavior of $N$-point functions in their soft limits~\cite{Baumann:2011nk}. Moreover, the perturbative coupling between the observable sector and the isocurvature sector leads to the boost in the collapsed limit relative to the squeezed limit.

\vskip 6pt
The outline of the paper is as follows: In Section~\ref{sec:general}, we present new proofs of various important theorems~\cite{Suyama:2007bg, Maldacena:2002vr} concerning the soft limits of primordial correlation functions.  Although these results aren't new, some of our methods and perspectives may be. In particular, we make a bit more precise the analogy between soft limits of inflationary correlation functions and soft pion limits of QCD~\cite{Adler:1964um, Weinberg:1966kf, Weinberg:1967kj, Weinberg:1996kr, Coleman, Donoghue:1992dd}. In Appendix~\ref{sec:AppA}, we present a number of technical details and generalizations related to the theorems of Section~\ref{sec:general}.
 In Section~\ref{sec:QSFI}, we compute the soft limits of QSFI explicitly in terms of the fundamental parameters of the theory.
We conclude in Section~\ref{sec:Discussion} with a discussion of the prospects of using non-Gaussianity as a particle detector.

\section{Generalities: Soft Limits of Inflationary Correlation Functions}
\label{sec:general}

We begin with proofs of various important theorems~\cite{Suyama:2007bg,Maldacena:2002vr} concerning the soft limits of inflationary correlation functions.

\subsection{Soft Internal Momenta}

We first consider the collapsed limit of the four-point function, i.e.~we let
 the internal momentum of the four-point function become soft, while keeping all external momenta hard. We will show that the four-point function in this limit is bounded from below by the squeezed limit of the corresponding three-point function. Our proof will make the role of additional degrees of freedom manifest. It will also make obvious how the proof generalizes to higher $N$-point functions.

\subsubsection{Suyama-Yamaguchi Bound}
\label{sec:Bound}

{\it Definitions.}---We characterize the soft limits of the three- and four-point functions by the following two quantities
\begin{align}
\hat{f}_{\mathsmaller{\rm NL}}&\equiv \frac{5}{12}\lim_{k_1 \to 0} \frac{\langle \zeta_{\k_1} \zeta_{\k_2} \zeta_{\k_3} \rangle'}{P_1 P_2} \ , \label{equ:def1}\\
\hat{\tau}_{\mathsmaller{\rm NL}}  &\equiv \frac{1}{4}\lim_{k_{12}\to 0} \frac{ \langle \zeta_{\k_1} \zeta_{\k_2} \zeta_{\k_3} \zeta_{\k_4} \rangle'}{P_1 P_3 P_{12}} \ , \label{equ:def2}
\end{align}
where $\langle \cdot \rangle'$ denotes the expectation value without the factor $(2\pi)^3 \delta(\sum \k_i)$. Moreover, we have defined $P_i \equiv P_\zeta(k_i) =\langle \zeta_{\k_i} \zeta_{\k_j}\rangle'$ and $k_{12} \equiv |\k_1+\k_2|$. All correlation functions are to be understood as `connected' correlation functions. For local non-Gaussianity, the definitions (\ref{equ:def1}) and (\ref{equ:def2}) reduce to the familiar (momentum-independent) definitions of $\fnl$~\cite{Komatsu:2001rj} and $\tnl$~\cite{Byrnes:2006vq}.
More generally, however, $\hat{f}_{\mathsmaller{\rm NL}}$ and $\hat{\tau}_{\mathsmaller{\rm NL}} $ can be momentum dependent---e.g.~in \S\ref{sec:QSFI}, we will find $ \hat{f}_{\mathsmaller{\rm NL}}(k_1) \propto k_1^{3- \alpha}$ and $\hat{\tau}_{\mathsmaller{\rm NL}}(k_{12}) \propto k_{12}^{6 - 2 \alpha}$, with $3\geq \alpha \geq \tfrac{3}{2}$.
We will now give a proof of the bound $\hat{\tau}_{\mathsmaller{\rm NL}} \geq (\tfrac{6}{5} \hat{f}_{\mathsmaller{\rm NL}})^2$~\cite{Suyama:2007bg, Smith:2011if}. In fact, we will give a proof of a slightly more general inequality, which includes $\hat{\tau}_{\mathsmaller{\rm NL}} \geq (\tfrac{6}{5} \hat{f}_{\mathsmaller{\rm NL}})^2$ as a special case.

\vskip 4pt
\noindent
{\it Collapsed four-point function.}---We start by considering the following four-point function,\footnote{Our conclusions will be unchanged if we replace $ \zeta^2(\x, t_\star) \to  \zeta(\x, t_\star) \zeta(\x+ {\vec{\varepsilon}}, t_\star)$ and $ \zeta^2(\0, t_\star) \to  \zeta(\0, t_\star) \zeta( {\vec{\varepsilon}}, t_\star)$, which may be useful in order to avoid divergences (or contact terms) when the operators are at coincident points.}
\beq
\langle  \zeta^2(\x, t_\star) \zeta^2 (\0, t_\star) \rangle \equiv \bra{\Omega} \zeta^2(\x, t_\star) \zeta^2 (\0, t_\star) \ket{\Omega} \ .
\eeq where $t_\star$ is some time at which $\zeta$ has become frozen  and $\ket{\Omega}$ is the vacuum\footnote{It is worth remarking that our results hold even if $\ket{\Omega}$ is not the vacuum state, as long as the state is invariant under spatial translations and rotations.} of the interacting theory.
Next, we insert a complete set of momentum eigenstates $\ket{n_\q}$ in the correlation function, \begin{align}
\langle  \zeta^2(\x, t_\star) \zeta^2 (\0, t_\star) \rangle &= \sum_{n ,\q} \, \langle  \zeta^2 (\x,  t_\star)| n_\q \rangle \langle  n_\q\hskip 1pt | \zeta^2({\0},  t_\star) \rangle \ , \label{equ:3}\end{align}
where $ \sum_{n ,\q}\, (\cdot) \equiv \sum_n \int_\q\, (\cdot) = \sum_n \int \frac{\d^3 \q}{(2\pi)^3} (\cdot)$.  We are working in the Heisenberg picture and therefore the states are formally\footnote{In practice, we define the Hilbert space perturbatively using the free theory.  The states then become implicitly time dependent through the projection onto the interacting vacuum.} independent of time.  Here, $\q$ is the total momentum of a state and $n$ is a label for individual states 
(e.g.~`particle number'\hskip 1pt\footnote{Of course, the notion of `particle states' is subtle in time-dependent cosmological backgrounds without asymptotically free states. None of those subtleties will affect our arguments, so we will, for convenience, sometimes (ab)use the particle physics terminology. \label{foot:particles}} states or different particle species) within the momentum eigenspace. In the following, we will drop the time label $t_\star$. Using $\langle \zeta^2(\x) \ket{n_\q} = e^{i \q \cdot \x} \langle \zeta^2({\0}) \ket{n_\q} $, we can write
eq.~(\ref{equ:3}) as
\beq
\langle \zeta^2(\x) \zeta^2(\0)\rangle_{\k} \, \equiv \, \int \d^3 x \, e^{-i {\k} \cdot {\x}} \langle \zeta^2 ({\x}) \zeta^2({\0}) \rangle =  \sum_n  |\langle n_\k  | \zeta^2({\0}) \rangle |^2  \ . \label{equ:15new}
\eeq
Notice that the r.h.s.~is written as a sum over positive definite terms.  If we can show that one of the contributions is the square of the squeezed limit of the three-point function of $\zeta$, then we can conclude that the collapsed four-point function---the l.h.s.~of (\ref{equ:15new})---is always greater or equal.

\vskip 4pt
\noindent
{\it Squeezed three-point function.}---To complete the proof, we insert a complete set of states in the three-point function $\langle \zeta(\x) \zeta^2({\0})  \rangle$, i.e.
\beq
\langle  \zeta_\k| \zeta^2({\0}) \rangle = \int \d^3 x \, e^{+ i {\k} \cdot {\x}} \langle \zeta(\x) \zeta^2({\0})  \rangle  = \sum_{n, \q}\, \langle \zeta_\k | n_\q \rangle \langle n_\q \hskip 1pt | \zeta^2 ({\0}) \rangle \ , \label{equ:14}
\eeq
where $\bra{\zeta_\k} =\ket{\zeta_\k}^\dagger = \bra{\Omega} \zeta_{-\k}$.
It is always possible to choose the
basis states $\ket{n_\q}$, such that 
\beq
\langle \zeta_\k  | n_\q \rangle = (2\pi)^3 \delta({\k}-{\q} \hskip 1pt)\, \delta_{1 n} \, P_\zeta^{1/2}(k) \ .  \label{equ:15}
\eeq
	\begin{table}[h!]

	\heavyrulewidth=.08em
	\lightrulewidth=.05em
	\cmidrulewidth=.03em
	\belowrulesep=.65ex
	\belowbottomsep=0pt
	\aboverulesep=.4ex
	\abovetopsep=0pt
	\cmidrulesep=\doublerulesep
	\cmidrulekern=.5em
	\defaultaddspace=.5em
	\renewcommand{\arraystretch}{1.6}

	\begin{center}
		\small
		\begin{tabular}{ll}
			\toprule 
				$\ket{\Omega}$&  vacuum \\[-5pt] 
		$\ket{1_\k} \propto \ket{\zeta_\k} $ &  ``one-zeta"
				\\[-5pt]
				$\ket{\tilde n_\k} \perp \ket{\zeta_\k}$ &  ``everything else"
				\\[4pt]	
 			\bottomrule
		\end{tabular}
	\end{center}
	\vspace{-0.5cm}
	\caption{Spectrum of states and their meanings. 
	\label{table:spectrum}}
	\end{table}
	\noindent
Up to an arbitrary phase, this implies
$| 1_\k\rangle = P_\zeta^{-1/2}(k)\, | \zeta_\k \rangle$.
In an abuse of the standard particle physics terminology, we will call $\ket{1_\k}$ the `one-zeta' state.
All other states, which we will denote by~$\ket{\tilde n_\k}$, are orthogonal to $\ket{\zeta_\k}$. They may be associated with `single-particle' states corresponding to additional degrees of freedom or simply `multi-zeta' states, i.e.~states created with higher powers of the operator $\zeta$.
Substituting (\ref{equ:15}) into~(\ref{equ:14}), we get
\beq
\langle \zeta_\k | \zeta^2({\0}) \rangle = \langle 1_\k |  \zeta^2 ({\0}) \rangle\, P_\zeta^{1/2}(k) \ . \label{equ:9new}
\eeq

\vskip 4pt
\noindent
{\it Bound on the collapsed four-point function.}---We can therefore write eq.~(\ref{equ:15new}) as
\beq
\fbox{$\displaystyle \langle \zeta^2(\x) \zeta^2(\0)\rangle_{\k} \, =\, \frac{|\bra{\zeta_{\k}} \zeta^2(\0) \rangle|^2}{P_\zeta(k)} \, +\, \sum_{\tilde n} |\bra{ \tilde n_{\k} } \zeta^2(\0) \rangle|^2 $}\ . \label{equ:result1}
\eeq
Since $\displaystyle \sum_{\tilde n }  |\langle \tilde n_\k |  \zeta^2({\0})  \rangle |^2 \ \geq\ 0$, we conclude that
\beq\label{equ:inequality}
 \langle \zeta^2(\x) \zeta^2(\0)\rangle_{\k} \, \ge\, \frac{|\bra{\zeta_{\k}} \zeta^2(\0) \rangle|^2}{P_\zeta(k)} \ . 
\eeq
If we take the soft limit $k \to 0$, we get a relation between the collapsed four-point function and the squeezed three-point function. 
Moreover, from our derivation it is easy to see how the result can be generalized to give a relation between the
soft limits of $2N$-point functions and the squeezed limits of $(N+1)$-point functions,
\beq
\langle \zeta^N(\x) \zeta^N(\0)\rangle_{\k} \, =\, \frac{|\bra{\zeta_{\k}} \zeta^N(\0) \rangle|^2}{P_\zeta(k)} \, +\, \sum_{\tilde n} |\bra{ \tilde n_{\k} } \zeta^N(\0) \rangle|^2 \ .
\eeq
\vskip 4pt
\noindent
{\it Suyama-Yamaguchi bound.}---Eq.~(\ref{equ:inequality}) can be written as
\beq
\int \limits_{\q_1} \int \limits_{\q_2} \langle \zeta_{\q_1} \zeta_{\k - \q_1} \zeta_{\q_2} \zeta_{-\q_2 - \k} \rangle'  \ \ge\ \frac{\left| \int_{\q_1} \langle \zeta_\k \hskip 1pt \zeta_{\q_1} \zeta_{-\q_1 - \k}\rangle' \right|^2}{P_\zeta(k)}  \ , 
\eeq
which, in the limit $k \to 0$, becomes 
\beq
\int \limits_{\q_1} \int \limits_{\q_2} \hat \tau_{\mathsmaller{\rm NL}} \,P_\zeta(q_1) P_\zeta(q_2)  \ \ge \ \int \limits_{\q_1} \int \limits_{\q_2}( \tfrac{6}{5} \hat f_{\mathsmaller{\rm NL}})^2 \, P_\zeta(q_1) P_\zeta(q_2) \ .
\eeq
So far this is completely general. To get the specific form of the Suyama-Yamaguchi bound~\cite{Suyama:2007bg} one has to make a further assumption.
It is often the case that $\hat \tau_{\mathsmaller{\rm NL}}$ and $(\hat f_{\mathsmaller{\rm NL}})^2$ are either momentum independent or have the same momentum dependence (so that we can still extract a momentum-independent amplitude). 
We then get 
\beq
\fbox{$\displaystyle \hat{\tau}_{\mathsmaller{\rm NL}} \geq (\tfrac{6}{5} \hat{f}_{\mathsmaller{\rm NL}})^2 $}\ . \label{equ:QED}
\eeq

\subsubsection{Interpretation}

What is the physical meaning of the extra contributions in eq.~(\ref{equ:result1})? It is well-known \cite{Byrnes:2006vq, Suyama:2007bg} that curvature fluctuations arising from a `single source' (which may or may not be the inflaton) saturate the Suyama-Yamaguchi inequality (at least at tree level). In this case, the squeezed three-point function of $\zeta$ completely determines the collapsed four-point function---i.e.~the extra contributions in (\ref{equ:result1}) 
are subdominant, in the sense that
\beq
\displaystyle \lim_{k \to 0} \frac{\bra{\tilde n_\k} \zeta^2 \rangle}{\bra{1_\k} \zeta^2 \rangle} = 0 \qquad ({\rm single\ source\ at\ tree\ level})\ . \label{equ:SingleSource}
\eeq
Non-negligible $\bra{\tilde n_{\k \to 0}} \zeta^2 \rangle$ can have the following origins:
\begin{enumerate}

\item {\it Extra fields.} A straightforward way to have 
contributions from extra states
is to have multiple sources for the primordial fluctuations.
This can lead to a significant boost of the collapsed four-point function  relative to the square of the squeezed three-point function, $\hat{\tau}_{\mathsmaller{\rm NL}} \gg (\tfrac{6}{5} \hat{f}_{\mathsmaller{\rm NL}})^2$.
As a measure of such `multi-source' non-Gaussianity we can define
\beq
X \, \equiv\, \lim_{k \to 0} \left[ \frac{\langle \zeta^2(\x) \zeta^2(\0)\rangle_{\k} \, P_\zeta(k) }{|\bra{\zeta_{\k}} \zeta^2(\0) \rangle|^2} - 1\, \right] \, =\,  \lim_{k \to 0}\, \sum_{\tilde n} \frac{|\bra{ \tilde n_{\k} } \zeta^2(\0) \rangle|^2}{|\bra{ 1_{\k} } \zeta^2(\0) \rangle|^2}  \ . \label{equ:result1c}
\eeq
We will see an explicit example of this situation in Section~\ref{sec:QSFI}.

\item {\it Gravitons.} Strictly speaking, even single-source inflation always has additional 
states corresponding to the graviton, i.e.~tensor metric perturbations. In fact, in single-field {\it slow-roll} inflation these states give the dominant contribution to the collapsed four-point function, $\hat  \tau_{\mathsmaller{\rm NL}} \sim {\cal O}(\epsilon)$~\cite{Seery:2008ax}, where $\epsilon \equiv - \dot H/H^2 < 1$. 
This is parametrically larger than $\hat{f}_{\mathsmaller{\rm NL}}^2 \sim {\cal O}(\epsilon^2)$~\cite{Maldacena:2002vr}, although both contributions are too small to be observable. Being interested in observable signatures, we will typically suppress the graviton contributions in our results. It would be easy to include their effects as ${\cal O}(\epsilon)$ corrections.

\item {\it Excited initial states.}  The effects of excited initial states on primordial non-Gaussianity have been explored in~\cite{Chen:2006nt,Holman:2007na, Agullo:2010ws, Ganc:2011dy, Agullo:2011aa, Kundu:2011sg}.  In particular, it was pointed out~\cite{Agullo:2010ws} that non-trivial contributions to soft limits are possible. It is conceivable that these effects are captured by $\bra{\tilde n_{\k\to 0}} \zeta^2 \rangle$
in our formalism, but we leave a detailed analysis for the future~\cite{Excited}.

\item {\it Loops.} Even in single-source inflation we can (formally) have $\hat{\tau}_{\mathsmaller{\rm NL}} > (\tfrac{6}{5} \hat{f}_{\mathsmaller{\rm NL}})^2$ from loops~\cite{Suyama:2010uj, Byrnes, Sugiyama:2012tr}. 
In single-clock inflation~\cite{Cheung:2007st}, these loop contributions aren't physical and can be removed by the right choice of physical coordinates~\cite{Senatore:2009cf}. 
In more general single-source scenarios, loops may contribute non-trivially.
Having said that, in this paper, we won't consider these loop effects, but instead focus on tree-level contributions. 
\end{enumerate}

\subsection{Soft External Momenta} \label{sec:softex}

We now use similar ideas to derive the famous consistency relations for correlation functions with soft external momenta~\cite{Maldacena:2002vr,Creminelli:2004yq}.  Our approach will draw inspiration from methods used in proving the low-energy theorems of theories with spontaneous symmetry breaking~\cite{Weinberg:1996kr,Coleman}, such as the classic soft pion theorems of QCD~\cite{Adler:1964um, Weinberg:1966kf, Weinberg:1967kj}. This connection was emphasized recently in~\cite{Hinterbichler:2012nm}.
The proofs in this section will be restricted to the limit $k \to 0$, but we will show in Appendix~\ref{sec:AppA} how they can be generalized to finite~$k$.

\subsubsection{Symmetries of Inflation}

The consistency relations arise from the symmetries of inflation, which we therefore quickly review (for further discussion see~\cite{Maldacena:2011nz, Creminelli:2011mw,Hinterbichler:2012nm}).

\begin{figure}[h!]
   \centering
      \hspace{-1.6cm} \includegraphics[scale =0.45]{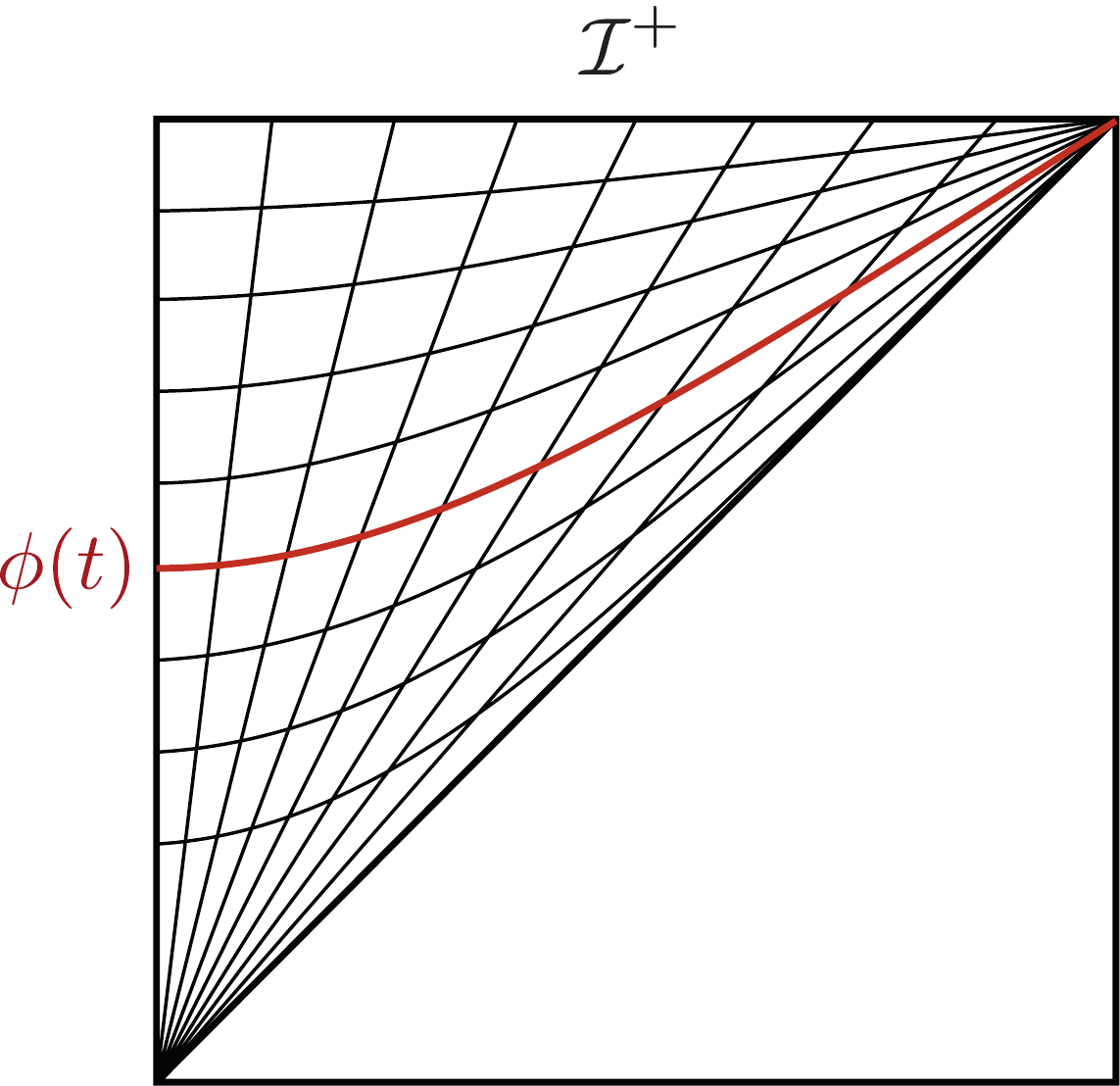}
   \caption{Conformal diagram of de Sitter space. The inflationary background picks out a preferred time-slicing.}
  \label{fig:dS}
\end{figure}

\vskip 4pt
Inflation is characterized by a quasi-de Sitter background
\beq
\d s^2 \simeq - \d t^2 + e^{2Ht}\hskip 1pt \d \x\hskip 1pt{}^2\ .
\eeq
The isometry group of de Sitter space, $SO(4,1)$, contains 6 translations and rotations, 1 dilation
\beq
Ht \ \mapsto\ H t -  \lambda \ ,\qquad \x \ \mapsto\ (1+\lambda) \x\ , 
\eeq
and 3 special conformal transformations (SCTs)
 \beq
Ht \ \mapsto\ Ht - 2 \hskip 1pt \vec{b} \cdot \x \ , \qquad \x \ \mapsto\ \x + \vec{b}(H^{-2} e^{-2Ht} - \x\hskip 1pt{}^2) + 2 (\vec{b} \cdot \x) \hskip 1pt \x\ ,
\eeq
 where $\lambda$ and $\vec{b}$ are infinitesimal parameters. At late times, $t \to \infty$, these isometries act as conformal transformations on the spatial boundary ${\cal I}^+$:
 \begin{align}
 \x &\ \mapsto \ (1+\lambda) \x \ , \label{equ:d}\\
 \x &\ \mapsto\ \x - \vec{b} \hskip 2pt \x\hskip 1pt{}^2 + 2 (\vec{b} \cdot \x) \hskip 1pt \x\ . \label{equ:sct}
 \end{align}
 The time-dependence of the inflationary background $\phi(t)$ spontaneously breaks the symmetry under dilations and special conformal transformations.
 Inflationary correlations are therefore, in general, not invariant under the conformal transformations.\footnote{The symmetries are still explicit for fields that don't couple to the inflaton, such as tensor metric fluctuations~\cite{Maldacena:2011nz} or scalar spectator fields~\cite{Creminelli:2011mw, Kehagias:2012pd}. Even for inflaton fluctuations, the symmetries are restored in the so-called decoupling limit, where $M_{\rm pl} \to \infty$, $\dot H \to 0$, with $M_{\rm pl}^2 \dot H \to const.$}  However, since the symmetry is non-linearly realized, we still expect non-trivial relations between different $N$-point functions.
 
 \vskip 4pt
As explained nicely in recent work by Creminelli, Nore\~na, and Simonovi\'c~\cite{Creminelli:2012ed} and
Hinterbichler, Hui and Khoury~\cite{Hinterbichler:2012nm}, the symmetries in eqs.~(\ref{equ:d}) and (\ref{equ:sct}), in fact, have significance beyond quasi-de Sitter backgrounds. They are symmetries of adiabatic modes in any FRW background.\footnote{Adiabatic modes are defined locally as coordinate redefinitions of the unperturbed FRW solution. Coordinate transformations of the form (\ref{equ:d}) and (\ref{equ:sct}) with time-dependent parameters $\lambda(t)$ and $\vec{b}(t)$ induce the following co-moving curvature perturbation
\beq
\zeta = \lambda(t) + 2\hskip 1pt \vec{b}(t) \cdot \x  \ . \label{equ:local}
\eeq
For {\it constant} transformation parameters $\lambda$ and $\vec{b}$ this gauge mode is the $k \to 0$ limit of a physical solution~\cite{Creminelli:2012ed, Hinterbichler:2012nm}, a la Weinberg~\cite{Weinberg:2003sw}, with suitable fall-off at spatial infinity.}
The Goldstone boson associated with the spontaneous breaking of the symmetry is the curvature perturbation\footnote{For spontaneously broken `spacetime' symmetries the number of Goldstone bosons does not have to equal the number of broken generators~\cite{Low:2001bw}.} $\zeta$, which in co-moving gauge is defined by~\cite{Bardeen:1983qw, Salopek:1990jq}
\beq
g_{ij}(\x,t) = a^2(t) e^{2 \zeta(\x,t)}\delta_{ij} \ . \label{equ:zeta}
\eeq
As we will see, the squeezed limit of $N$-point functions of $\zeta$ are related to the variation of $(N-1)$-point functions under dilations and special conformation transformations. This is a consequence of the conformal transformations being non-linearly realized.

\vskip 4pt
We now show how these considerations can be incorporated into our formalism. 
For concreteness, we focus on the {dilation symmetry}\footnote{It is our understanding that a complete treatment will appear in~\cite{Hinterbichler, Goldberger} (see also \cite{Creminelli:2012ed}).} in co-moving coordinates
\begin{align}
\x &\ \mapsto\ (1+\lambda) \x\ , \\
\zeta &\ \mapsto\ \zeta + \lambda ( 1 + \x \cdot \partial_\x\hskip 1pt \zeta ) \ .
\end{align}
  This symmetry is a large\footnote{A ``large diffeomorphism" is one that does not vanish at infinity.  These transformations are distinct from the gauge redundancies of the theory, namely local diffeomorphisms that vanish at infinity.} diffeomorphism that is non-linearly realized by the ``gauge-invariant" perturbation $\zeta$.  It should not be confused with an isometry of FRW (which is linearly realized) or a residual\footnote{There is a residual gauge transformation in ``$\zeta$-gauge", eq.~(\ref{equ:zeta}), that does not act on $\zeta$~\cite{Weinberg:2008zzc}.} gauge transformation.  During inflation, only time diffeomorphisms are broken by the background~\cite{Cheung:2007st}.  Therefore, this large spatial diffeomorphism will act linearly on all the non-gravitational degrees of freedom.
Associated with the dilation symmetry of the action is a conserved current $J^\mu_d$ and a charge $Q_d \equiv \int \d^3 x \hskip 1pt  J^0_d (\x,t)$. 
Although the charge does not exist  after symmetry breaking (due to an IR divergence), local commutators are still well defined, e.g.
\begin{align}
\delta_d\hskip 1pt \zeta &\equiv i [Q_d, \zeta] = -1- \x \cdot \partial_\x \hskip 1pt  \zeta\ . \label{equ:13}
\end{align}

For future reference, we will need the matrix element $\bra{\Omega} Q_d \hskip 1pt \zeta_\q \hskip 1pt \ket{\Omega} \equiv \langle Q_d \ket{\zeta_\q}$.
Invariance of the vacuum under spatial translations and rotations, fixes the form of this matrix element to be
\beq
\langle Q_d \ket{\zeta_\q} = i f(q) \cdot (2\pi)^3\delta(\q \hskip 1pt)\ ,
\label{eq:matrixJpi}
\eeq
where $f$ is a function that depends only on the magnitude of the 3-momentum $q \equiv |\q \hskip 1pt|$. 
In order for this to be consistent with the commutator (\ref{equ:13}), we require ${\rm Re}[f(q)] = \tfrac{1}{2}$.
 The imaginary part of $f(q)$ won't contribute in what we are about to say, so we can set it to zero.
 We then have
 \begin{align}
  \langle Q_d \ket{\zeta_\q} &=  \tfrac{i}{2} \cdot (2\pi)^3 \delta(\q \hskip 1pt)\ .
 \end{align}
 
\subsubsection{Consistency Relations}

 We now use these results to give a quick proof of the `Adler zero'~\cite{Adler:1964um} for the squeezed limit of the three-point function in single-field inflation~\cite{Maldacena:2002vr, Creminelli:2004yq}.  
 We will also see how extra fields have to enter in order to produce a non-trivial signal in the squeezed limit.
 
 \vskip 4pt
 Associated with the conserved current $J^\mu_d$ is the following Ward identity~\cite{Weinberg:1995mt}
 \begin{align}
i\, \partial_\mu^{(x)} \langle J^{\mu}_d(\x,t) \zeta(\y, t_\star) \zeta(\z, t_\star,) \rangle &\ =\ \delta(t-t_\star)\delta({\x}-{\y} \hskip 1pt ) \hskip 2pt \langle \delta_d  \hskip 1pt\zeta({\y}, t_\star) \, \zeta({\z}, t_\star) \rangle \nonumber\\ &\hspace{1cm} +\, \delta(t-t_\star) \delta({\x}-{\z} \hskip 1pt) \, \langle   \zeta({\y}, t_\star)\hskip 2pt \delta_d  \hskip 1pt\zeta({\z}, t_\star) \rangle  \ , \label{equ:Ward0}
\end{align}
where for clarity we have this time shown all the time dependences explicitly. 
 If we integrate\footnote{The integral over time is extended in the imaginary direction because our correlation functions are ``time ordered" with respect to the integration contour in the complex plane (see Appendix~\ref{sec:AppA} for further details). } eq.~(\ref{equ:Ward0}) over $\int^{t_\star +i\epsilon}_{t_\star -i\epsilon} \d t \int \d^3 x$, we arrive at
 \begin{align}
  \langle [Q_d, \zeta_{\k_1} \hskip -1pt \zeta_{\k_2}] \rangle  &\, =\,   \langle [Q_d, \zeta_{\k_1}] \zeta_{\k_2} \rangle  +  \langle \zeta_{\k_1} [Q_d, \zeta_{\k_2}] \rangle\ , \label{equ:Ward}
 \end{align}
 where we have reverted to dropping the common time coordinate $t_\star$.
  In going from (\ref{equ:Ward0}) to (\ref{equ:Ward}), we have dropped the integral over $\partial_i J^i_d$ because it only receives contributions from spatial infinity (see Appendix~\ref{sec:AppA}).
  With the help of (\ref{equ:13}), we can express the r.h.s.~of (\ref{equ:Ward}) as
  \begin{align}
 \langle [Q_d, \zeta_{\k_1} \hskip -1pt \zeta_{\k_2}] \rangle  &=   -i(3 + \k_1 \cdot \partial_{\k_1} + \k_2 \cdot \partial_{\k_2} ) \langle \zeta_{\k_1} \hskip -1pt \zeta_{\k_2} \rangle' \, (2\pi)^3 \delta(\k_1+\k_2) \ . \label{equ:1}
 \end{align}
 Here and in the following, we only keep terms that contribute to the connected part of the correlation functions.
 Next, we insert a complete set of states in the l.h.s.~of (\ref{equ:Ward}),
\begin{align}
\langle [Q_d, \zeta_{\k_1} \hskip -1pt \zeta_{\k_2}] \rangle & \ =\  \sum_{n,\q} \Big[\langle Q_d \ket{n_\q} \bra{n_\q \hskip 1pt} \zeta_{\k_1} \hskip -1pt \zeta_{\k_2} \rangle  - \langle \zeta_{\k_1} \hskip -1pt \zeta_{\k_2} \ket{n_\q} \bra{n_\q \hskip 1pt} Q_d \rangle \Big] \ ,
\end{align}
where 
\beq
\langle Q_d \ket{n_\q} \equiv \tfrac{1}{2}c_n(q)\cdot (2\pi)^3 \delta(\q \hskip 1pt)\ .
\eeq 
Since $c_0 = \vev{Q_d}' \equiv \bra{\Omega} Q_d \ket{\Omega}' \in \mathbb{R}$, the vacuum insertion doesn't contribute. 
We separate the `one-zeta' state, with $c_1(q) = \tfrac{i}{2}  P_\zeta^{-1/2}(q)$ (up to a removable phase), 
from the rest\footnote{The functions $|c_{\tilde n}(q)|^2$ in eq.~(\ref{equ:2}) are analogous to the spectral density functions $\rho(q^2)$ in the K\"all\'en-Lehmann representation of correlation functions~\cite{Weinberg:1996kr}.  In principle, one could determine the properties of $|c_{\tilde n}(q)|^2$ from~$\langle J^\mu_d(\x) J^\nu_d(\0) \rangle$.} 
\begin{align}
\langle [Q_d, \zeta_{\k_1} \hskip -1pt \zeta_{\k_2}] \rangle 
&\ = \  \lim_{q \to 0} \left[ \frac{i}{P_\zeta(q) }  \bra{\zeta_\q \hskip 1pt}
 \zeta_{\k_1} \hskip -1pt \zeta_{\k_2} \rangle' + i\sum_{\tilde n} {\rm Im} \Big[ c_{\tilde n}(q) \bra{\tilde n_\q\hskip 1pt } \zeta_{\k_1} \hskip -1pt \zeta_{\k_2} \rangle' \Big] \right] (2\pi)^3\delta(\k_1+\k_2) \ . \label{equ:2}
 \end{align}
 In deriving eq.~(\ref{equ:2}) we have used the fact that 
\begin{align}
 \langle\zeta_{\k_1} \hskip -1pt \zeta_{\k_2}|\tilde{n}_\q \rangle = \langle \tilde{n}_\q \hskip 1pt |\zeta_{-\k_1} \hskip -1pt \zeta_{-\k_2} \rangle^*= \langle \tilde{n}_{-\q} \hskip 1pt|\zeta_{\k_1}\hskip -1pt \zeta_{\k_2} \rangle^* \ ,
\end{align}
 where the second equality comes from the invariance of correlation functions under parity. 
 Comparing eqs.~(\ref{equ:1}) and (\ref{equ:2}), we get 
 \begin{align}
\fbox{$ \displaystyle  \frac{\bra{\zeta_\q \hskip 1pt} \zeta_{\k_1} \hskip -1pt \zeta_{\k_2} \rangle'}{P_\zeta(q)}  \ \xrightarrow{q\to 0} \  - \Big[ 3 + \k_1\cdot \partial_{\k_1} + \k_2\cdot \partial_{\k_2}   \Big] \langle \zeta_{\k_1} \hskip -1pt \zeta_{\k_2} \rangle'   - \sum_{\tilde n } {\rm Im} \Big[ c_{\tilde n}(q) \bra{\tilde n_\q \hskip 1pt} \zeta_{\k_1} \hskip -1pt \zeta_{\k_2} \rangle' \Big]  $}
\ . \label{equ:zero}
\end{align}
This result generalizes straightforwardly to the squeezed limits of higher $N$-point functions (see Fig.~\ref{fig:soft1})
\begin{align}
\frac{\bra{\zeta_\q \hskip 1pt} \zeta_{\k_1} \cdots \zeta_{\k_N} \rangle'}{P_\zeta(q)}  &\ \xrightarrow{q\to 0} \  -   \Big[ 3(N-1) + \sum_{i=1}^N \k_i \cdot \partial_{\k_i}  \Big] \langle \zeta_{\k_1} \cdots \zeta_{\k_N} \rangle'   \nonumber \\ &  \hspace{3.5cm}-\ \sum_{\tilde n} {\rm Im} \Big[ c_{\tilde n}(q) \bra{\tilde n_\q \hskip 1pt} \zeta_{\k_1} \cdots \zeta_{\k_N} \rangle' \Big] 
\ . \label{equ:zeroN}
\end{align}
Moreover, it is easy to see how this result implies consistency relations for soft internal momenta~\cite{Senatore:2012wy} (see Fig.~\ref{fig:soft2}),
\begin{align}
\frac{\langle ( \zeta_{\k_1} \dots \zeta_{\k_M}) ( \zeta_{\k_{M+1}} \dots \zeta_{\k_{N}}) \rangle'}{P_\zeta(q)} &\ \xrightarrow{q\to 0} \ \frac{\langle \zeta_{\k_1} \dots \zeta_{\k_N} \ket{\zeta_\q}'}{P_\zeta(q)}  \frac{\bra{\zeta_\q \hskip 1pt} \zeta_{\k_{M+1}} \dots \zeta_{\k_{N}}\rangle'}{P_\zeta(q)}   \nonumber \\ &  \hspace{1.3cm}-\, \frac{1}{P_\zeta(q)} \sum_{\tilde n} \langle  \zeta_{\k_1} \dots \zeta_{\k_M} \ket{\tilde n_\q}' \bra{\tilde n_\q \hskip 1pt}  \zeta_{\k_{M+1}} \dots \zeta_{\k_{N}} \rangle' \ , \label{equ:zeroNM}
\end{align}
where $\q \equiv \sum_{i=1}^M \k_i$. Again, this was derived by inserting a complete set of states and separating the `one-$zeta$' state $\ket{1_\q} = P_\zeta^{-1/2}(q) \ket{\zeta_\q}$ from all states orthogonal to it, $\ket{\tilde n_\q}$.

\subsubsection{Interpretation}

There are two important consequences of the $q \to 0$ limit.  First, in this limit  causality explains the absence of a potential boundary term arising from the $\partial_i J^i_d$ term in (\ref{equ:Ward0}) (see Appendix~\ref{sec:AppA} for the complete argument). Second,  in this limit the `multi-particle' states among $\ket{\tilde n_\q}$ are not expected to contribute the leading divergent term, so that we can drop them on the r.h.s.~of~(\ref{equ:zero}).  It is straightforward to check this in most examples, however, a general proof\,\footnote{Even in the context of S-matrix elements for the scattering of Goldstone bosons, there are known exceptions to this statement~\cite{Weinberg:1996kr} (\S19.2). } is beyond the scope of this paper.

\vskip 4pt
In the special case of single-field inflation, 
eq.~(\ref{equ:zero}) therefore reduces to the famous single-field consistency relation~\cite{Maldacena:2002vr}\hskip 1pt\footnote{Excited initial states may contribute corrections at finite $q$, or when the number of $e$-folds is taken to infinity before taking $q \to 0$ (see e.g.~\cite{Holman:2007na,Agullo:2010ws,Ganc:2011dy,Agullo:2011aa,Kundu:2011sg}). This is related to the observation that $q \ll k_{1,2}$ is not a sufficient condition for satisfying the consistency relation (see e.g.~\cite{Flauger:2010ja, Chen:2010xka, Creminelli:2011rh}).  Understanding these finite $q$ effects can be important given that observations will only probe a finite range of momenta.}
\begin{align}
\frac{ \langle\zeta_\q \hskip 1pt | \zeta_{\k_1} \hskip -1pt \zeta_{\k_2} \rangle' }{P_\zeta(q)} &\ \xrightarrow{q\to 0} \    -\Big[ 3 + \k_1\cdot \partial_{\k_1} + \k_2\cdot \partial_{\k_2}   \Big] \langle \zeta_{\k_1} \hskip -1pt \zeta_{\k_2} \rangle'  \ =\ (1-n_s) \hskip 1pt P_\zeta(k_1) \ . \label{equ:zero2}
\end{align}
For scale-invariant fluctuations,\footnote{Scale-invariance is natural if the theory has an internal shift symmetry, $\phi \to \phi + c$. Both this symmetry and the `spacetime' dilation symmetry of de Sitter space are spontaneously broken by the time-dependent inflationary background, $\phi(t)$. However, a linear combination of shift and spacetime dilation is preserved~\cite{Nicolis:2011pv}. At late times, this diagonal symmetry becomes the spatial dilation symmetry that we discuss in this section. This explains the scale-invariance of the two-point function~\cite{Creminelli:2010ba}. Softly breaking the shift symmetry allows for weakly scale-dependent fluctuation spectra.} i.e.~$\langle \zeta_{\k_1} \hskip -1pt \zeta_{\k_2} \rangle' = (k_1 k_2)^{-3/2}$ or\, $n_s=1$, the r.h.s.~of (\ref{equ:zero2}) is zero and the squeezed limit of the three-point function vanishes.   
This is analogous to the Adler zero of soft pion physics~\cite{Adler:1964um} (see also \cite{Weinberg:1996kr, Coleman}).

In multi-field inflation, we can get contributions to the squeezed limit, even for scale-invariant fluctuations,
\begin{align}
\frac{ \langle \zeta_\q\hskip 1pt | \zeta_{\k_1}\hskip -1pt \zeta_{\k_2} \rangle'}{P_\zeta(q)}  &\ \xrightarrow{q\to 0} \  -\sum_{\tilde n} {\rm Im} \Big[ c_{\tilde n}(q) \bra{\tilde n_\q \hskip 1pt} \zeta_{\k_1} \hskip -1pt \zeta_{\k_2} \rangle' \Big]  \ . \label{equ:zero3}
\end{align}
This expression doesn't just tell us that we need extra fields to generate a squeezed limit, but also how the extra fields must enter. 
We will study an explicit example of this situation in Section~\ref{sec:QSFI}.

\section{Case Study: Soft Limits of Quasi-Single-Field Inflation}
\label{sec:QSFI}

In this section, we will illustrate some of the abstract results of the previous section with the concrete example of quasi-single-field inflation (QSFI)~\cite{Chen:2009zp}.  This is a particularly interesting case study, since QSFI exhibits non-trivial behaviour in its soft limits: 
\begin{enumerate}
\item[i)] the soft limits of $N$-point functions have characteristic momentum scalings that depend on the mass of the isocurvaton field; 
\item[ ii)] the collapsed limits of $2N$-point functions are naturally boosted. 
\end{enumerate}
In \S\ref{sec:review}, we explain these features of QSFI qualitatively.  In \S\ref{sec:explicit}, we compute the soft limits of the three- and four-point functions explicitly.

\subsection{Review of Quasi-Single-Field Inflation}
\label{sec:review}

\begin{figure}[h!]
   \centering
       \includegraphics[scale =0.5]{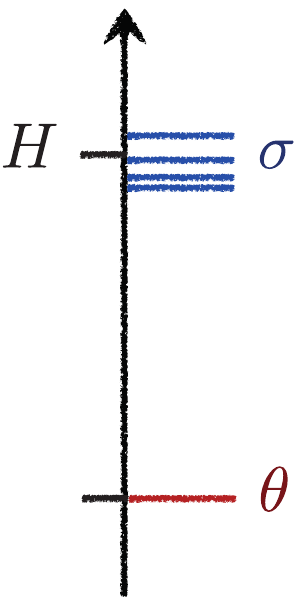}
   \caption{Particle spectrum of quasi-single-field inflation.}
  \label{fig:QSFI}
\end{figure}

\noindent
{\it Dynamics.}---QSFI couples a massive `isocurvaton' field $\sigma$ to an `adiabatic' perturbation $\theta$ (see Fig.~\ref{fig:QSFI}). The dynamics of these fields is governed at leading order by the following free field Hamiltonian~(density)
\beq
 {\cal H}_0 = \tfrac{1}{2} (\partial_\mu \theta)^2 + \tfrac{1}{2}(\partial_\mu \sigma)^2 + \tfrac{1}{2} m^2 \sigma^2 \ , \label{equ:S0}
\eeq
where $m \sim H$.
At linear order, the adiabatic perturbation $\theta$ and the curvature perturbation $\zeta$ are related by
\beq
\theta = - \sqrt{2\epsilon}\hskip 1pt \Mp\, \zeta\ ,
\eeq
where $\epsilon \equiv - \dot H/H^2$.
A quadratic mixing term, e.g.~${\cal H}_{\rm mix} = - \rho\hskip 1pt  \dot \theta \sigma$, converts $\sigma$ into $\theta$ (or $\zeta$). This can communicate large interactions in the isocurvaton sector to the observable sector. Our basic example will be the cubic interaction ${\cal H}_3 =  \mu \sigma^3$.
Following~\cite{Chen:2009zp}, we treat ${\cal H}_{\rm mix}$ as part of the interaction Hamiltonian
\beq
{\cal H}_{\rm int} =  - \rho \hskip 1pt  \dot \theta \sigma  + \mu \sigma^3\ . \label{equ:Lint}
\eeq
In order to describe the mixing perturbatively we require $\rho < H$.

\begin{figure}[h!]
   \centering
       \includegraphics[scale =0.29]{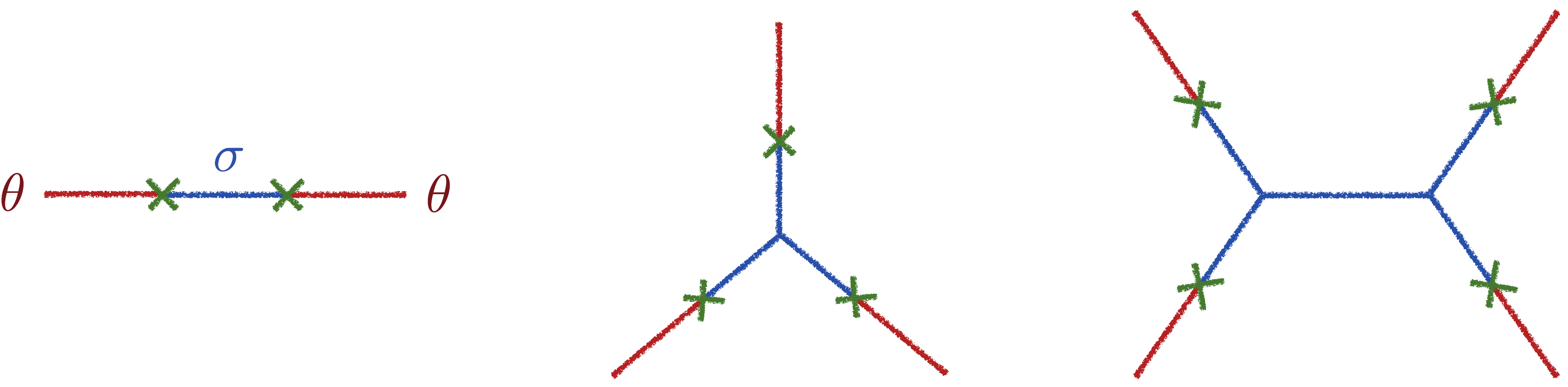}
   \caption{Perturbative description of $N$-point functions in quasi-single-field inflation.}
  \label{fig:Feynman}
\end{figure}

\vskip 4pt
\noindent
{\it Quantization.}---The fields are quantized in the interaction picture
\begin{align}
\hat \theta_\k(\tau) &= u_k(\tau) \hat a_\k + u_k^*(\tau) \hat a_{-\k}^\dagger \ ,  \\
\hat \sigma_\k(\tau) &= v_k(\tau) \hat b_\k + v_k^*(\tau) \hat b_{-\k}^\dagger \ , 
\end{align}
where $\d \tau = \d t /a(t)$ is conformal time and $u_k(\tau)$ and $v_k(\tau)$ are the free-field mode functions
\begin{align}
u_k(\tau) &= \frac{H}{\sqrt{2k^3}}(1+ik\tau) e^{-ik\tau} \ , \label{equ:u}\\
v_k(\tau) &= \sqrt{\frac{\pi}{2}} \frac{H}{\sqrt{2k^3}} (-k\tau)^{3/2} \, \H_\nu^{\mathsmaller{ (1)}}(-k\tau)\ ,\ \quad\text{with}~~\nu \equiv \sqrt{\frac{9}{4}- \frac{m^2}{H^2}}\ . \label{equ:v}
\end{align}
Here, $ \H_\nu^{\mathsmaller{ (1)}}$ is the Hankel function of the first kind.
The massive mode decays on superhorizon scales, $v_k(\tau) \sim (-\tau)^{3/2-\nu}$.
This feature completely explains the non-trivial scaling of correlation functions in the soft limits~\cite{Baumann:2011nk}.
The mode functions have been normalized such that $[\hat a_\k, \hat a^\dagger_{\k'}] = [\hat b_\k, \hat b^\dagger_{\k'}] = (2\pi)^3 \delta(\k+\k')$, which makes $\hat a_\k$ ($\hat a_\k^\dagger$) and $\hat b_\k$ ($\hat b_\k^\dagger$) the standard annihilation (creation) operators. Since there is no $\theta$-$\sigma$ coupling in ${\cal H}_0$, we also have $[\hat a_\k, \hat b_\k^\dagger] = 0$.
We compute correlation functions of products of field operators, $Q(\tau)$, in the $in$-$in$ formalism~\cite{Weinberg:2005vy,Chen:2010xka}
\beq
\vev{Q(\tau)} = \langle \Omega | \left[ \bar{\rm T} \exp \left(i \int_{-\infty}^\tau \d \tau' \,  \hat H_{\rm int}(\tau')  \right) \right]  \hat Q(\tau)  \left[ {\rm T} \exp \left(- i \int_{-\infty}^\tau \d \tau' \,  \hat H_{\rm int}(\tau')  \right) \right] | \Omega \rangle \ , \label{equ:master}
\eeq
where the symbols ${\rm T}$ and $\bar{\rm T}$ denote time-ordering and anti-time-ordering, respectively. The interaction Hamiltonian $H_{\rm int} =  \int \d^3 x \, a^4\,{\cal H}_{\rm int}$ can be written as a sum of two terms 
\beq
H_2 \equiv - \int\d^3x\, a^3 \rho\hskip 1pt \theta'\sigma\qquad {\rm and} \qquad
H_3 \equiv \int\d^3x\,a^4 \mu\sigma^3\ .  \label{equ:H}
\eeq
The mixing term $H_2$ is treated as a perturbative insertion in Feynman diagrams (see Fig.~\ref{fig:Feynman}).

\vskip 4pt
\newpage
\noindent
{\it Soft limits.}---We end this review section with a brief summary of the soft limits of QSFI:
\begin{itemize}
\item The three-point function has the following momentum scaling in the squeezed limit
\beq
\lim_{k_1 \to 0} \frac{ \langle \zeta_{\k_1} \zeta_{\k_2} \zeta_{\k_3} \rangle'}{P_\zeta(k_1)} \ \propto \ k_1^{\frac{3}{2} - \nu} \ . \label{equ:sq}
\eeq
This result has a simple explanation~\cite{Baumann:2011nk}:
The long (or soft) $\sigma$-mode exits the horizon at the time $ \tau_1 \sim k_1^{-1}$, after which it decays with a rate that depends on its mass, $\sigma_k(\tau) \sim (-\tau)^{3/2-\nu}$.  Its correlation with the short (or hard) modes is evaluated at the horizon crossing of the short modes, i.e.~at $\tau_2 \sim k_2^{-1}$. By this time, the amplitude of the long $\sigma$-mode has decayed by a factor of $(\tau_2/\tau_1)^{3/2-\nu} = (k_1/k_2)^{3/2-\nu}$. This explains the suppression of the squeezed limit in eq.~(\ref{equ:sq}). Below we will confirm this intuition with an exact calculation.

\item The four-point function in the collapsed limit scales as
\beq
\lim_{k_{12} \to 0} \frac{ \langle  \zeta_{\k_1} \zeta_{\k_2} \zeta_{\k_3} \zeta_{\k_4 }\rangle' }{P_\zeta(k_{12})}\ \propto \ k_{12}^{3 - 2\nu} \ . \label{equ:Scaling4}
\eeq
Again, the scaling follows from the superhorizon evolution of the soft $\sigma$-mode.

\item The amplitudes of the collapsed four-point function and the squeezed three-point function are related as
\beq
\hat \tau_{\mathsmaller{\rm NL}} \ \sim\ \frac{(\tfrac{6}{5} \hat f_{\mathsmaller{\rm NL}})^2}{\left( \frac{\rho}{H}\right)^2} \ \gg \  (\tfrac{6}{5} \hat f_{\mathsmaller{\rm NL}})^2\ .
\eeq
We can understand this from the Feynman diagrams in Fig.~\ref{fig:Feynman}. While in the three-point function every $\sigma$-propagator attaches to an external $\theta$ (or $\zeta$), in the four-point function the virtual $\sigma$-mode does not have to be converted to $\zeta$. The four-point function therefore only gets four powers of the small coupling $\rho/H$, while the square of the three-point function  has six powers of $\rho/H$.
For fixed three-point amplitude, we therefore get a boost in the four-point amplitude by a factor of $(\rho/H)^{-2} \gg 1$.
\end{itemize}

\subsection{Explicit `in-in' Computations}
\label{sec:explicit}

We now want to go beyond the qualitative discussion of the previous section, and compute the soft limits of the three- and four-point functions explicitly in terms of the fundamental parameters of the theory.

\subsubsection{Power Spectrum}

To assess the regime of validity of perturbation theory, we first use the $in$-$in$ master formula (\ref{equ:master}) to compute the leading correction to the power spectrum of the adiabatic mode $\theta$,
\beq
 \vev{\theta_{\k_1}\theta_{\k_2}} = i^2\int_{-\infty}^0\d\tau_1\int_{-\infty}^{\tau_1}\d\tau_2\,\vev{[ \hat H_2(\tau_2),[ \hat H_2(\tau_1), \hat \theta_{\k_1} \hat\theta_{\k_2}]]} \ .
\eeq
Substituting eqs.~(\ref{equ:u}), (\ref{equ:v}) and (\ref{equ:H}), we find~\cite{Chen:2009zp}
\beq\label{eqn:qsfipower}
  \Delta_\zeta^2 \equiv k^3 P_\zeta(k) = \frac{1}{4} \frac{H^4}{ \Mp^2 |\dot H|} \left[ 1 + c(\nu) \left( \frac{\rho}{H} \right)^2 + \cdots \right]\ ,
\eeq
where
\begin{align}
c(\nu) &\equiv 2\pi \, {\rm Re} \Big[ \int_0^\infty \d x_1 \int_{x_1}^\infty \d x_2 \left( x_1^{-1/2} \H_\nu^{\mathsmaller{(1)}}(x_1) e^{i x_1} x_2^{-1/2} \H_\nu^{\mathsmaller{(2)}}(x_2) e^{-i x_2} \right.  \nonumber \\
& \hspace{4.7cm} \left.  -\, x_1^{-1/2} \H_\nu^{\mathsmaller{(1)}}(x_1) e^{-i x_1} x_2^{-1/2} \H_\nu^{\mathsmaller{(2)}}(x_2) e^{-i x_2} \right)  \Big] \ , \label{equ:cnu} \\
&=\ \ \ \ \includegraphicsbox[scale=.37]{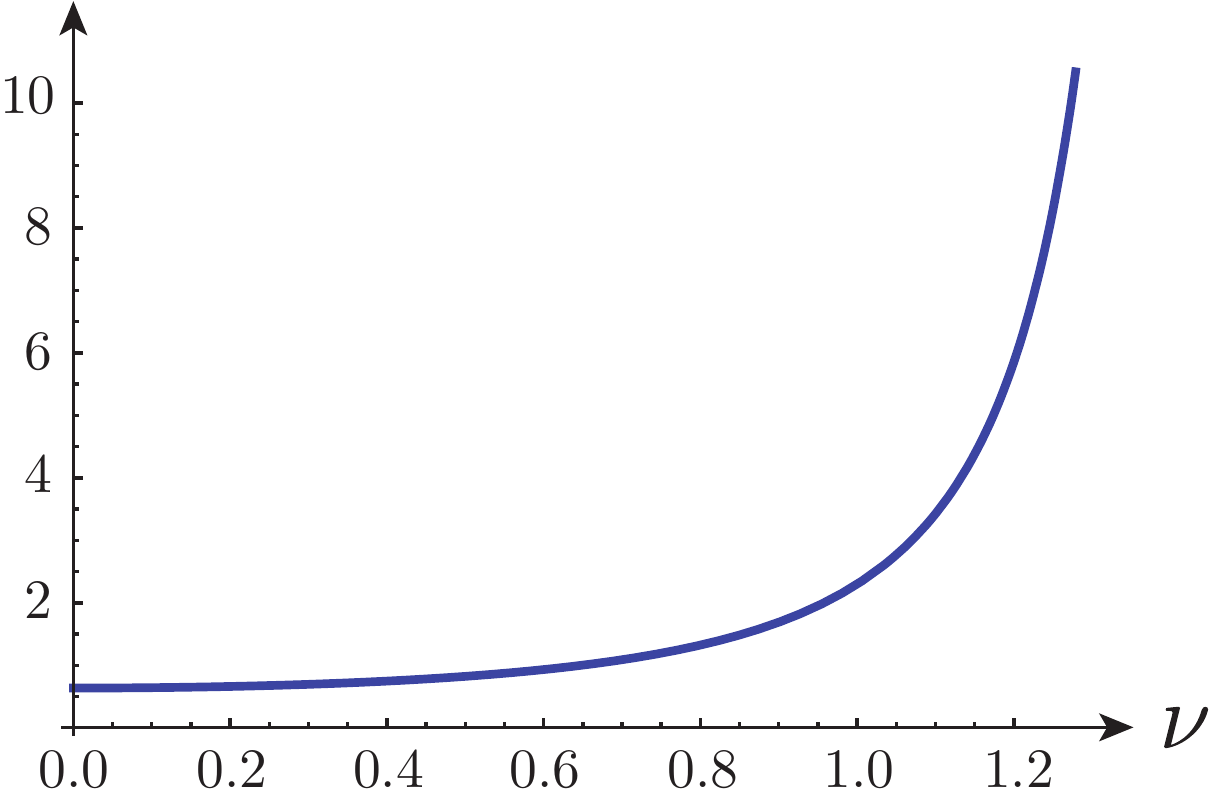} \nonumber
\end{align}

\vskip 6pt
\noindent
The function $c(\nu)$ formally diverges in the limit $\nu \to \frac{3}{2}$. In this limit, the isocurvaton $\sigma$ becomes massless and freezes after horizon exit, instead of decaying. As a result, the transfer of $\sigma$ to $\zeta$ continues for an indefinite time after horizon exit and evaluating the curvature perturbation at the infinite future $\tau=0$ is no longer justified. Instead, the bispectrum should be evaluated at the time $\tau_{f}$ when inflation ends. This regulates the divergence.\footnote{For the specific mixing term we consider,  ${\cal H}_{\rm mix} = - \rho \hskip 1pt  \dot \theta \sigma$, an effective mass of order $\rho$ is generated for $\sigma$ \cite{Chen:2009zp, Baumann:2011su}.  If we were to re-sum the perturbation series, we expect that the divergence to be regulated by the effective mass.  However, it isn't guaranteed that this possibility exists for generic mixing interactions, so we will ignore it here as well.}
Finally, we remark that the calculation is under perturbative control if and only if
\beq
c(\nu) \left(\frac{\rho}{H} \right)^2 < 1\ . \label{equ:control}
\eeq

\subsubsection{Soft Limits: Methodology}

We now derive the soft limits of QSFI in a style that will be familiar to cosmologists~\cite{Creminelli:2004yq}.
At the end of \S\ref{sec:bi} and \S\ref{sec:tri}, we will demonstrate the precise equivalence between this method and the formalism of \S\ref{sec:general}.

\vskip 4pt
\noindent
{\it Long--short split.}---In cosmology, soft limits are described as the correlation between long and short modes.
Following~\cite{Maldacena:2002vr, Creminelli:2004yq, Komatsu:2010}, we treat the long-wavelength fields as {\it classical} backgrounds for the short-wavelength {\it quantum} fields. The fields $\theta$ and $\sigma$ are split into long and short modes,\hskip 1pt\footnote{This is morally similar to the split into soft and hard modes in the factorization theorems of QCD~\cite{Collins:1989gx}.}
\beq
\theta = \tl + \ts \quad {\rm and} \quad \sigma = \sl + \ss,
\eeq
where
\beq
\tl(\x,\tau)=\int_{k < \Lambda} \dfrac{\d^3 k}{(2\pi)^3}\, \theta_{\k}(\tau)e^{i\k.\x} \qquad \text{and}\qquad \ts(\x,\tau)=\int_{k > \Lambda} \dfrac{\d^3 k}{(2\pi)^3} \, \theta_{\k}(\tau)e^{i\k.\x}~,
\eeq
and similarly for $\sigma$. The precise value of the momentum cut-off $\Lambda$ won't be important.
We then write the interaction Hamiltonian $H_{\rm int}$ in terms of $\tl$, $\ts$, $\sl$ and $\ss$. The quadratic mixing term $H_2$ doesn't produce any coupling between long and short modes. The only relevance of the mixing is therefore to convert $\ss$ into $\ts$ via 
\begin{align}
H_{\SS\SS} &= - \rho\hskip 1pt a^3 \int \d^3 x\, \ss \ts' \ . \label{equ:HSS}
\end{align}
The cubic interaction $H_3$, on the other hand, does couple long modes to short modes.
The only interaction that will be relevant in the soft limit is
\begin{align}
H_{\L \SS\SS } &= 3 \mu \hskip 1pt a^4 \int \d^3 x \, \sl \ss^2 \ . \label{equ:HLSS}
\end{align}

\vskip 4pt
\noindent
{\it Soft limits.}---Our goal is to use eqs.~(\ref{equ:HSS}) and (\ref{equ:HLSS})  to compute the squeezed limit of the three-point function,
\beq
\langle \theta_{\k_1} \theta_{\k_2} \theta_{\k_3} \rangle \xrightarrow{k_1\to 0}   \langle \theta_{\k_1} \langle \theta_{\k_2} \theta_{\k_3} \rangle_{\sl}  \rangle\ ,
\eeq
and the collapsed limit of the four-point function,
\beq
\langle \theta_{\k_1} \theta_{\k_2} \theta_{\k_3} \theta_{\k_4}\rangle \xrightarrow{k_{12}\to 0}   \langle \langle \theta_{\k_1} \theta_{\k_2} \rangle_{\sl}   \langle \theta_{\k_3} \theta_{\k_4} \rangle_{\sl} \rangle\ .
\eeq
We see that both soft limits can be expressed in terms of the modulated power spectrum
\beq
\langle \theta_{\k_1} \theta_{\k_2} \rangle_{\sl}\ , \label{equ:mp}
\eeq
i.e.~the power spectrum $\langle \theta_{\k_1} \theta_{\k_2} \rangle$ in the presence of a long-wavelength background $\sl$.
We compute eq.~(\ref{equ:mp}) in the $in$-$in$ formalism 
\beq
\vev{\theta_{\k_1}\theta_{\k_2}}_{\sigma_\L(\tau)} = i^3\int_{-\infty}^\tau \d\tau_1\int_{-\infty}^{\tau_1}\d\tau_2\int_{-\infty}^{\tau_2}\d\tau_3\;\vev{\left[\hat H_{\text{int}}(\tau_3),\left[\hat H_{\text{int}}(\tau_2),\left[\hat H_{\text{int}}(\tau_1), \hat \theta_{{\k_1}} \hat \theta_{{\k_2}}(\tau)\right]\right]\right]} \ ,
\eeq
where $H_{\rm int} = H_{\L \SS \SS} + H_{\SS \SS}$.  
In the limit $\tau \to 0$, this becomes 
\beq
\vev{\theta_{\k_1} \theta_{\k_2} }_{\sigma_\L} = -\dfrac{3\pi^2}{2} \frac{\mu \rho^2}{H^2} \frac{1}{k_1^3}\left(A_{12}+B_{12}\right)\ ,
\label{eq:vevtt}
\eeq
where the $A$-term is obtained by replacing $H_\text{int}(\tau_3)$ by $H_{\L \SS \SS}$ and the two other interaction terms by $H_{\SS \SS}$, while the $B$-term is obtained by replacing $H_\text{int}(\tau_2)$ by $H_{\L \SS \SS}$ and the rest by $H_{\SS \SS}$. This leads to 
\bea
A_{12}&\equiv&\int_{-\infty}^0 \d x_1\int_{-\infty}^{x_1} \d x_2\int_{-\infty}^{x_2} \d x_3 \; f_A(x_1,x_2,x_3)\; (\sigma_{\L})_{{\k_1}+{\k_2}} ({x_3/k_1}) \ ,\label{def:A}\\
B_{12}&\equiv&\int_{-\infty}^0 \d x_1\int_{-\infty}^{x_1} \d x_2\int_{-\infty}^{x_2} \d x_3\; f_B(x_1,x_2,x_3) \; (\sigma_{\L})_{{\k_1}+{\k_2}} ({x_2/k_1})\ , \label{def:B}
\eea
with
\begin{align}
f_A(x_i) & \equiv (-x_1 x_2 (x_3)^2)^{-1/2} \sin(-x_1)\sin(-x_2) \times\,\text{Im}\Big(\H_\nu^{\mathsmaller{(2)}}(-x_1)\, \H_\nu^{\mathsmaller{(2)}}(-x_2)(\H_\nu^{\mathsmaller{(1)}}(-x_3))^2\Big) \ ,  \nonumber \\
f_B(x_i) &\equiv (-x_1 (x_2)^2 x_3)^{-1/2} \sin(-x_1) \times\,\text{Im}\Big(\H_\nu^{\mathsmaller{(1)}}(-x_1)\, \H_\nu^{\mathsmaller{(2)}}(-x_2)\Big)\text{Im}\Big(\H_\nu^{\mathsmaller{(1)}}(-x_2) \, \H_\nu^{\mathsmaller{(2)}}(-x_3)e^{ix_3}\Big) \ . \nonumber
\end{align}

\subsubsection{Bispectrum in the Squeezed Limit}
\label{sec:bi}

With this preliminary work, it is now straightforward to compute the squeezed limit of the bispectrum,
\beq
\vev{\theta_{\k_1} \vev{\theta_{\k_2} \theta_{\k_3} }_{\sigma_\L} } = -\dfrac{3\pi^2}{2} \frac{\mu \rho^2}{H^2} \frac{1}{k_2^3}\Big( \vev{ \theta_{\k_1} A_{23}} + \vev{\theta_{\k_1} B_{23} }\Big)\ . \label{equ:bi}
\eeq
 The only slight complication is due to the fact that the calculation requires the unequal time correlation function of two fields, i.e.~$\vev{\sigma_\k(\tau)\theta_\q(0)}$. However, since $\theta_\q(\tau)$ freezes in the limit $| q \tau| \ll 1$, we have $\displaystyle \lim_{|q \tau| \ll 1}\theta_\q(\tau)\approx\theta_\q(0)$ and hence
\beq
\vev{\sigma_\k(\tau)\theta_\q(0)} \ \xrightarrow{|k\tau| \ll 1} \  \vev{\sigma_\k(\tau)\theta_\q(\tau)}\ ,
\eeq 
where
\bea
\vev{\sigma_{\k}(\tau)\theta_{\q}(\tau)} &=& - i\int_{-\infty}^\tau\d\tilde{\tau}\int_\p \, a(\tilde{\tau})^3 \rho\, \bra{0} \hat \sigma_{\p}(\tilde{\tau})\hat \theta_{-\p}'(\tilde{\tau}) \hat \sigma_{\k}(\tau) \hat \theta_{\q}(\tau)\ket{0} \ +\ c.c. \nonumber
\\
&=& -2 \rho\,(2\pi)^3\delta(\k+\q\,)\, \text{Re}\left[i\int_{-\infty}^\tau\d\tilde{\tau}\,a(\tilde{\tau})^3\,v_{k}(\tilde{\tau})v_{k}^*(\tau)u'_{q}(\tilde{\tau})u^*_{q}(\tau)\right]\ .
\eea
Using the superhorizon limits of the mode functions,
\beq
\lim_{|k \tau| \ll 1} u_{k}(\tau)\approx\dfrac{H}{\sqrt{2k^3}} \qquad \text{and} \qquad \lim_{|k \tau| \ll 1}  v_{k}(\tau)\approx -i\, \dfrac{2^\nu\Gamma(\nu)}{2\sqrt{\pi}}\dfrac{H}{\sqrt{k^3}}(-k\tau)^{3/2-\nu}  \ ,
\label{eq:shlimit}
\eeq
we get 
\beq
\vev{\sigma_\k(\tau) \theta_\q\hskip 1pt(\tau)} \ \xrightarrow{|k\tau| \ll 1} \ \sqrt{\frac{2}{\pi}}\frac{\Gamma(\nu)}{2^{2-\nu}} \frac{\rho}{H} \frac{H^2}{k^3} (-k \tau)^{3/2- \nu} \, d(\nu)\, (2\pi)^3 \delta(\p+\q\,)\ , \label{equ:2pt}
\eeq
where
\begin{align}
d(\nu) &\equiv \sqrt{\frac{\pi}{2}}\int_{-\infty}^0 \d x\, (-x)^{-1/2} \, {\rm Re} \left[ \H_\nu^{\mathsmaller{(1)}}(-x) e^{-i x}\right] \label{equ:dnu}  \ =\ \ \ \ \includegraphicsbox[scale=.4]{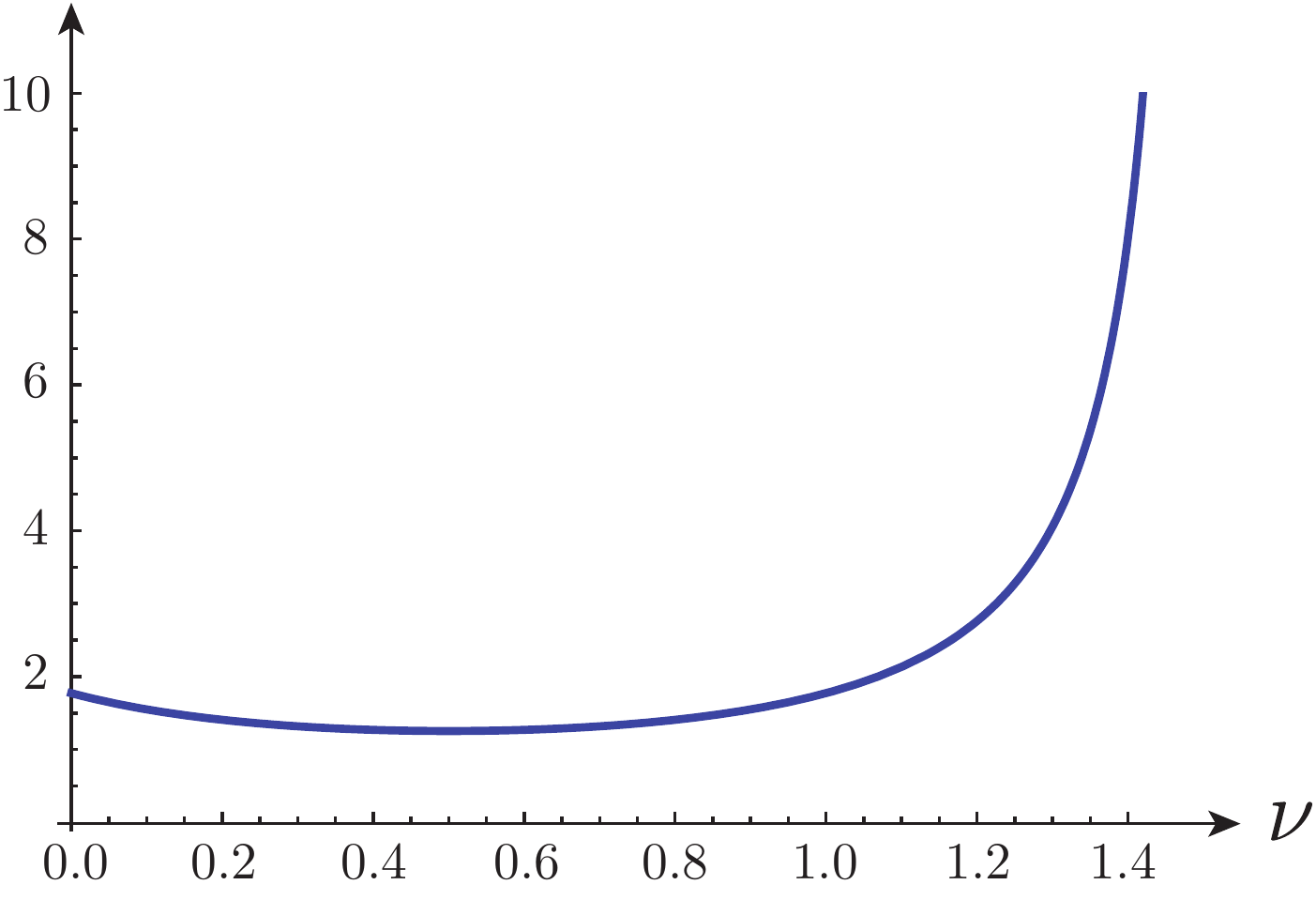}  
\end{align}
\vskip 6pt
\noindent
Substituting (\ref{equ:2pt}) into (\ref{equ:bi}), we obtain the squeezed limit of the bispectrum for the curvature perturbation $\zeta$,  
\beq
\vev{\zeta_{\k_1}\zeta_{\k_2} \zeta_{\k_3}}' \ \xrightarrow{k_1 \to 0} \ \frac{12}{5}\fnl(\nu) \cdot P_\zeta(k_1) P_\zeta(k_2) \left(\frac{k_1}{k_2} \right)^{3/2- \nu}  \ , \label{equ:Bfinal}
\eeq
where
\beq
\fnl(\nu) \equiv \, p(\nu) \, d(\nu) \cdot \frac{\mu \rho^3}{H^4} \cdot \Delta_\zeta^{-1}\ .
\eeq
Here, $d(\nu)$ is the same function as in eq.~(\ref{equ:dnu}) and $p(\nu)$ is defined as 
\beq
p(\nu) \equiv \, 5\pi^{3/2}\,\frac{\Gamma(\nu)}{2^{3-\nu}} \int_{-\infty}^0  \hskip -6pt\d x_1 \int_{-\infty}^{x_1} \hskip -3pt \d x_2\int_{-\infty}^{x_2} \hskip -3pt \d x_3 \; \Big[ (-x_3)^{3/2-\nu} f_A(x_i) +  (-x_2)^{3/2-\nu} f_B(x_i) \Big] \ .
\eeq
Eq.~(\ref{equ:Bfinal}) precisely matches the result that one obtains by taking the limit $k_1 \rightarrow 0$ of the full expression of the bispectrum \cite{Chen:2009zp}.

\vskip 10pt
\noindent
In the following insert we show how to reproduce eq.~(\ref{equ:Bfinal}) from our analysis in \S\ref{sec:general}:
\vskip 6pt
\hrule
\vskip 4pt
\noindent  
\small
We start by defining a complete set of one-particle states
\beq
\ket{1_\q^{\mathsmaller{(\zeta)}}} \equiv P_\zeta^{-1/2} \ket{\zeta_\q} \quad {\rm and} \quad \ket{1_\q^{\mathsmaller{(\tilde\sigma)}}} \equiv P_{\tilde \sigma}^{-1/2} \ket{\tilde \sigma_\q}\ . \label{equ:1states}
\eeq
In order for the states to be orthogonal, we have defined
\beq
\tilde \sigma_\q \equiv \sigma_\q -  \frac{\vev{\zeta_\q \hskip 1pt | \sigma_\q}'}{P_\zeta(q)}\, \zeta_\q\ .
\eeq
As our expansion parameter we define
\beq
 \varepsilon^2(\nu) \equiv  \frac{|\vev{\zeta \sigma}'|^2}{P_\zeta P_\sigma} < 1\ .
\eeq
To determine the effect of $\sigma$ on the squeezed limit of the three-point function of $\zeta$, we first  compute the parameter $c_{\tilde \sigma}$ in eq.~(\ref{equ:zero3}). Since spatial dilations act linearly on $\sigma$, we have
\beq
i \vev{[Q_d, \tilde \sigma_\q]} = \frac{\vev{\zeta_\q \hskip 1pt | \sigma_\q}'}{P_\zeta(q)} (2\pi)^3 \delta(\q\hskip 1pt)\ .
\eeq
This implies that
\beq
c_{\tilde \sigma}(q) = - i \frac{\vev{\zeta_\q \hskip 1pt |\sigma_\q}'}{P_\zeta P_{\tilde \sigma}^{1/2}}\ .
\eeq
The coefficient $c_{\tilde{\sigma}}$ may have a non-zero real part, but it won't contribute in what follows.
We therefore get
\begin{align}
{\rm Im}\left[c_{\tilde \sigma}(q)  \bra{1_\q^{\mathsmaller{(\tilde\sigma)}}} \zeta^2 \rangle' \right] &\ = \ - \frac{\bra{\zeta_\q \hskip 1pt} \sigma_\q \rangle'}{P_\zeta P_{\tilde \sigma}} \, {\rm Im}\left[i \bra{\tilde \sigma_\q \hskip 1pt} \zeta^2 \rangle' \right] \ \simeq\ - \frac{\bra{\zeta_\q \hskip 1pt} \sigma_\q \rangle' \bra{\sigma_\q \hskip 1pt} \zeta^2 \rangle'}{P_\zeta P_\sigma } \left[ 1 + {\cal O}(\varepsilon^2) \right]\ ,
\end{align}
and eq.~(\ref{equ:zero3}) becomes
\begin{align}
\frac{ \langle \zeta_\q \hskip 1pt \zeta_{\k_1} \hskip -1pt\zeta_{\k_2} \rangle'}{P_\zeta(q)}  &\ \xrightarrow{q\to 0} \  \frac{\bra{\zeta_\k} \sigma_\q \rangle' \bra{\sigma_\q\hskip 1pt} \zeta_{\k_1} \hskip -1pt \zeta_{\k_2} \rangle'}{P_\zeta(q) P_\sigma(q) }   \ .
\end{align}
This is equivalent to the starting point of this section, eq.~(\ref{equ:bi}), so following the same steps will lead to the answer in eq.~(\ref{equ:Bfinal}).

\vskip 4pt 
\hrule
\vskip 20pt
\normalsize

\newpage
\subsubsection{Trispectrum in the Collapsed Limit}
\label{sec:tri}

The collapsed limit of the four-point function is equal to the correlation of two modulated power spectra,
\beq
\vev{\vev{\theta_{{\k_1}}\theta_{{\k_2}}}_{\sigma_\L}\vev{\theta_{{\k_3}}\theta_{{\k_4}}}_{\sigma_\L}} = \left(\dfrac{3\pi^2}{2} \frac{\mu \rho^2}{H^2}\right)^2\dfrac{1}{k_1^3\,k_3^3}\Big(\underbrace{\vev{A_{12}A_{34}}+\vev{A_{12}B_{34}}+\vev{B_{12}A_{34}}+\vev{A_{12}B_{34}}}_{(\star)}\Big)\ , \label{equ:collapse}
\eeq
where the sum over contractions is
\begin{align}
(\star) &\ =\  \left(\, \prod_{i=1}^3 \int \d x_i \right) \left(\prod_{i=1}^3 \int \d y_i \right)  \times\  \\
&\hspace{0.7cm} \times\, \Big[\ f_A(x_i) f_A(y_i) \, \big \langle (\sl)_{\k_{12}}(\tfrac{x_3}{k_1})\, (\sl)_{\k_{34}}(\tfrac{y_3}{k_3}) \big \rangle \, + \, f_A(x_i) f_B(y_i)\, \big \langle (\sl)_{\k_{12}}(\tfrac{x_3}{k_1})\, (\sl)_{\k_{34}}(\tfrac{y_2}{k_3}) \big \rangle\nonumber\\
&\hspace{1.0cm} +\, f_B(x_i) f_A(y_i)\, \big \langle (\sl)_{\k_{12}}(\tfrac{x_2}{k_1})\, (\sl)_{\k_{34}}(\tfrac{y_3}{k_3}) \big \rangle  \, + \, f_B(x_i) f_B(y_i)\, \big \langle (\sl)_{\k_{12}}(\tfrac{x_2}{k_1})\, (\sl)_{\k_{34}}(\tfrac{y_2}{k_3}) \big \rangle \ \Big] \ . \nonumber
\end{align}
Using
\begin{align}
\vev{\sigma_{\q_1}(\tau_1)\sigma_{\q_2}(\tau_2)} &\ \xrightarrow{q_1,q_2\rightarrow 0}\ \dfrac{2^{2\nu-1}\Gamma(\nu)^2}{\pi}\dfrac{H^2}{2q_1^3}(-q_1\tau_1)^{3/2-\nu}(-q_2\tau_2)^{3/2-\nu}~(2\pi)^3\delta\left(\q_1+\q_2\right)\ ,
\end{align}
this becomes
\bea
(\star) &\ \to\ &\ \frac{\Gamma(\nu)^2}{2^{2-2\nu} \pi} \frac{H^2}{(k_1k_3)^{3/2-\nu}} \frac{1}{(k_{12})^{2\nu}} \, (2\pi)^3\, \delta \big({\k_1}+{\k_2}+{\k_3} + \k_4 \big) \times \nonumber \\
&& \hspace{0.5cm} \times\, \left[ \left( \prod_{i=1}^3 \int \d x_i \right) \Big(f_A(x_i)\, (-x_3)^{3/2-\nu} + f_B(x_i)\, (-x_2)^{3/2-\nu}\Big)\right]^2\ .
\eea
Eq.~(\ref{equ:collapse}) can therefore be written as 
\beq
\vev{\theta_{\k_1}\theta_{\k_2} \theta_{\k_3} \theta_{\k_4}}' \ \xrightarrow{k_{12}\to 0} \   t(\nu) \cdot \frac{\rho^4 \mu^2}{H^6} \frac{H^4}{(k_1 k_3)^{9/2- \nu}} \frac{1}{(k_{12})^{2\nu}}  \ , \label{equ:theta4}
\eeq
where 
\begin{align}
t(\nu) &\equiv  \left(\tfrac{6}{5}\,p(\nu)\right)^2  \ \ =\ \ \ \ \includegraphicsbox[scale=.45]{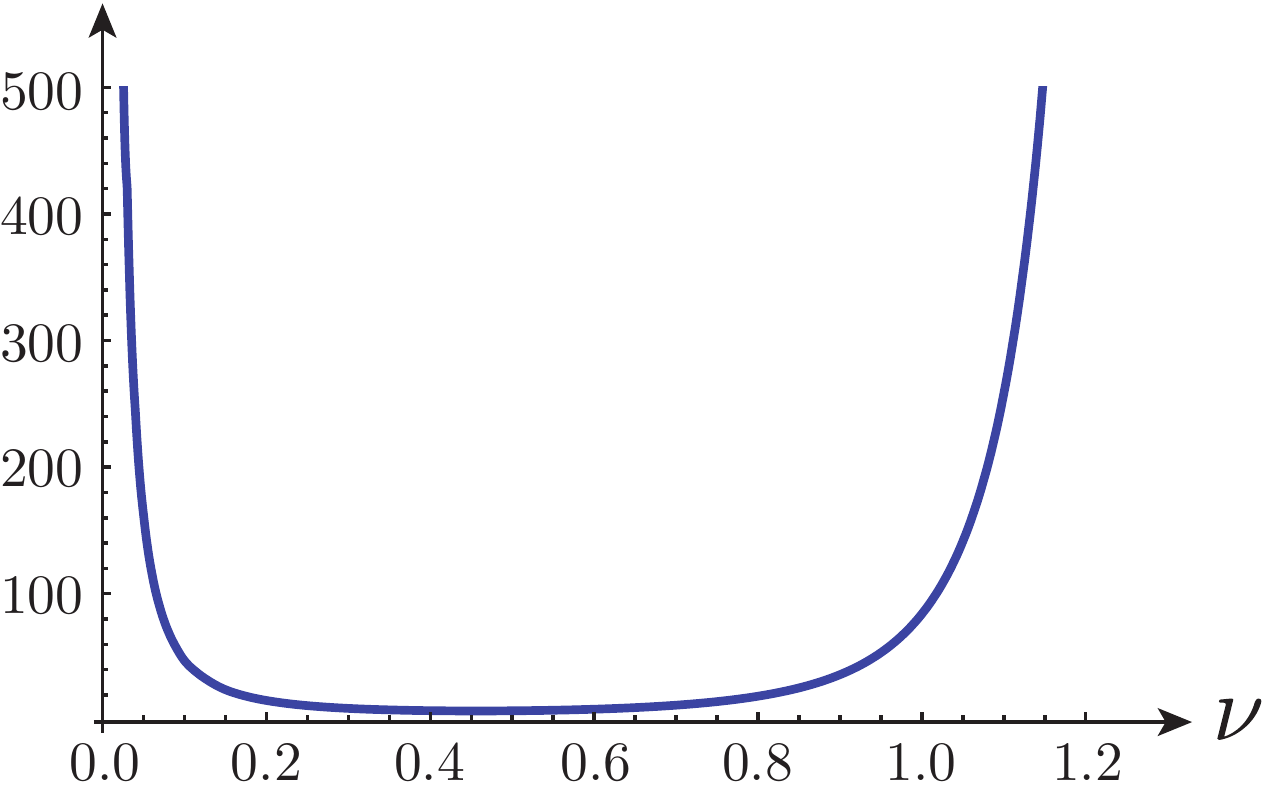} 
\end{align}
The function $t(\nu)$ diverges as $\nu\rightarrow0$ and $\nu\rightarrow \frac{3}{2}$. The divergence for $\nu\rightarrow 0$ arises from the fact that the expansion of the Hankel function in eq.~(\ref{eq:shlimit}) isn't justified when $k_{12}/k \ll e^{-1/\nu}$. Instead the expansion of the Hankel function should be changed to 
\beq
v_{k_{12}}(x/k_1) \rightarrow -i\sqrt{\frac{\pi}{2}}\dfrac{H}{\sqrt{2k_{12}^3}}\left(-\dfrac{k_{12}}{k_1}x\right)^{3/2}\ln\left(\frac{k_{12}}{k_1}\right)\ .
\eeq
This regulates the divergence.
The divergence of $t(\nu)$ for $\nu \to \frac{3}{2}$ has the same origin (and the same resolution) as the divergence of $c(\nu)$ in (\ref{equ:cnu}).
Finally, converting (\ref{equ:theta4}) to curvature perturbations, we get
\beq
\fbox{$\displaystyle \vev{\zeta_{\k_1}\zeta_{\k_2} \zeta_{\k_3} \zeta_{\k_4}}' \ \xrightarrow{k_{12}\to 0} \  4\, \tnl(\nu) \cdot P_\zeta(k_1) P_\zeta(k_3) P_\zeta(k_{12}) \left(\frac{k_{12}}{(k_1 k_3)^{1/2}} \right)^{3-2\nu}$} \ ,
\label{equ:Tfinal}
\eeq
where 
\beq
\fbox{$\displaystyle \tnl(\nu) =  \frac{(\tfrac{6}{5}\fnl(\nu) )^{2} }{\varepsilon^2(\nu)} $}\ , \qquad {\rm with} \qquad \varepsilon(\nu ) \equiv \frac{\rho}{H} d(\nu)\ .
\eeq
Note that we have recovered the scaling in eq.~(\ref{equ:Scaling4}).
Since $d^2(\nu) \simeq c(\nu)$, the condition on perturbative control, eq.~(\ref{equ:control}), becomes
 \beq
 \varepsilon(\nu) = \frac{\rho}{H} d(\nu) < 1\ .
 \eeq
 Hence, as long as the calculation is under perturbative control, we find $\tnl \ge (\tfrac{6}{5} \fnl)^2$, consistent with the
 Suyama-Yamaguchi inequality~\cite{Suyama:2007bg}.
 
\vskip 10pt
\noindent
Finally, we show how eq.~(\ref{equ:Tfinal}) also follows from our analysis in \S\ref{sec:general}:
\vskip 6pt
\hrule
\vskip 4pt
\noindent  
\small
  \noindent
We insert the complete set of one-particle states, eq.~(\ref{equ:1states}), into the four-point function
\beq
\langle\zeta^4\rangle = \int\limits_{\q} \left[ \langle\zeta^2|1_\q^{\mathsmaller{(\zeta)}}\rangle\langle1_\q^{\mathsmaller{(\zeta)}}|\zeta^2\rangle+\langle\zeta^2|1_\q^{\mathsmaller{(\tilde\sigma)}}\rangle\langle1_\q^{\mathsmaller{(\tilde\sigma)}}|\zeta^2\rangle \right] \ ,
\label{equ:4pt}
\eeq
where
\beq
\ket{1_\q^{\mathsmaller{(\tilde\sigma)}}} = N\left[\ket{1_\q^{\mathsmaller{(\sigma)}}} - \alpha\, \ket{1_\q^{\mathsmaller{(\zeta)}}} \right]\ ,
\eeq
with  $\alpha = \langle 1_\q^{\mathsmaller{(\zeta)}}|1_\q^{\mathsmaller{(\sigma)}} \rangle' \sim \varepsilon(\nu)$ and $N=1+ \mathcal{O}(\varepsilon)$.
At leading order in $\varepsilon$,
eq.~(\ref{equ:4pt}) becomes
\beq
\langle\zeta_{\k_1}\zeta_{\k_2}\zeta_{\k_3}\zeta_{\k_4}\rangle' \xrightarrow{k_{12}\rightarrow0}\frac{1}{P_\sigma(k_{12})}\langle\zeta_{\k_1}\zeta_{\k_2}  \sigma_{-\k_{12}}\rangle'\langle\sigma_{-\k_{34}}  \zeta_{\k_3}\zeta_{\k_4}\rangle' \ .
\eeq
This is equivalent to the starting point of this section,
eq.~(\ref{equ:collapse}), so following the same steps will lead to the answer in eq.~(\ref{equ:Tfinal}).

\vskip 4pt 
\hrule
\vskip 20pt
\normalsize

\newpage
\section{Discussion: Non-Gaussianity as a Particle Detector}
\label{sec:Discussion}

\begin{figure}[h!]
   \centering
       \includegraphics[scale =0.55]{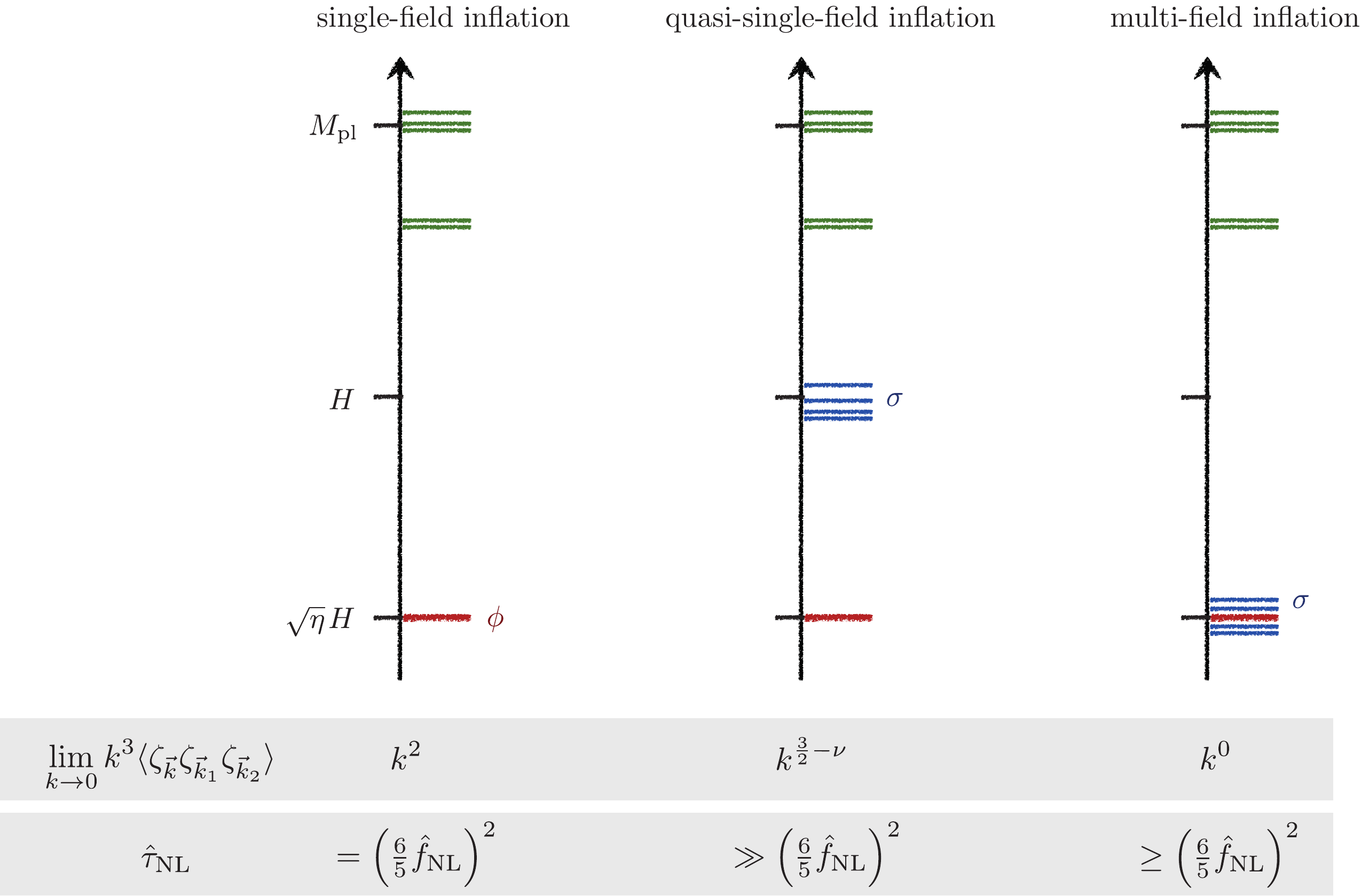}
   \caption{Particle spectra of inflationary models and their observational signatures.}
  \label{fig:spectrum}
\end{figure}

Observations of the CMB and LSS probe the physics of inflation through the statistics of the primordial curvature perturbations.
Soft limits of the correlation functions of these perturbations are especially sensitive to the number of light degrees of freedom during inflation. In this paper, we have explained how the spectrum of particles can impact observations (see Fig.~\ref{fig:spectrum}).

\vskip 4pt
\noindent
{\it Single-field inflation.}---The fluctuations in single-field inflation are Goldstone bosons of a spontaneously broken dilation symmetry~\cite{Hinterbichler:2012nm}. Being non-linearly realized, the broken symmetry is still reflected in Ward identities relating the squeezed limits of $N$-point functions to the symmetry variation of $(N-1)$-point functions~\cite{Maldacena:2011nz, Creminelli:2011mw, Hinterbichler:2012nm}. 
Up to small corrections associated with the breaking of scale-invariance, the squeezed limits of all $N$-point functions therefore vanish in single-field inflation~\cite{Maldacena:2002vr}.  This is analogous to the Adler zero~\cite{Adler:1964um} of chiral perturbation theory~\cite{Weinberg:1996kr}. In this note, we have given a new proof of this famous result.
Measuring a significant three-point function in the squeezed limit would therefore automatically rule out \/{\it all}\/ single-field models.
Going beyond single-field inflation, the squeezed limit captures information about the spectrum of additional light degrees of freedom.

\vskip 4pt
\noindent
{\it Multi-field inflation.}---Coupling additional fields to the inflaton relaxes some of the symmetry constraints on inflationary correlation functions. The extra fields can therefore produce a significant signal in the squeezed limit. In fact, if the non-Gaussianity arises from superhorizon evolution, then the signal tends to be peaked in the squeezed configuration.
If a single field (but not necessarily the inflaton) creates the primordial curvature perturbations $\zeta$ and their non-Gaussianity, then the collapsed limit of the four-point function is uniquely fixed by the squeezed limit of the three-point function. 
However, this correlation is broken if more than one field sources $\zeta$ and its interactions.
In that case one can show that the collapsed four-point function is always larger than the square of the squeezed three-point function~\cite{Suyama:2007bg}. In this note, we have given a new proof of this result.

\vskip 4pt
\noindent
{\it Quasi-single-field inflation.}---Supersymmetric models of inflation often mix a light inflaton field with an additional isocurvaton field whose mass is of order the inflationary Hubble scale~\cite{Baumann:2011nk}.
Such models of quasi-single-field inflation~\cite{Chen:2009zp} naturally give rise to a boosted four-point function and to characteristic momentum scalings in the soft limits. We have computed these signatures explicitly in terms of the fundamental parameters of the theory.  As we have explained elsewhere~\cite{Baumann:2011nk,stochastic}, these results are relevant for connecting the microscopic physics of QSFI to late-time observables such as stochastic halo bias~\cite{stochastic}. 

\vskip 6pt 
The above shows that soft limits of primordial non-Gaussianity provide a remarkable opportunity to probe the spectrum of particles during inflation, i.e.~at energies totally inaccessible to conventional particle physics experiments.
The PLANCK satellite~\cite{PLANCK}  
will measure the CMB anisotropies over a large range of scales and hence will significantly improve the observational constraints on the soft limits of primordial non-Gaussianity. 
At the same time, the Sloan Digital Sky Survey~\cite{SDSS} continues to collect data on the clustering of collapsed objects. The scale-dependent biasing~\cite{Dalal:2007cu} of these objects is a very sensitive probe of the squeezed limit of the primordial correlation functions~\cite{Schmidt:2010gw}.   Looking ahead, LSS-experiments like EUCLID~\cite{EUCLID} and LSST~\cite{LSST}, as well as CMB-experiments like PIXIE~\cite{PIXIE} and CMBPol~\cite{CMBPol} all have the potential to teach us a great deal about the soft limits of inflationary correlations.

\acknowledgments
Xingang Chen, Paolo Creminelli, Justin Khoury, Marko Simonovi\'c and Gianmassimo Tasinato kindly provided comments on a draft of the paper.  We thank Raphael Flauger and Rafael Porto for helpful discussions.
D.B.~and V.A.~gratefully acknowledge support from a Starting Grant of the European Research Council (ERC STG grant 279617). 
The research of D.G.~is supported by the DOE under grant number DE-FG02-90ER40542 and the Martin A.~and Helen Chooljian Membership at the Institute for Advanced Study.  
D.B.~thanks the theory groups at Groningen, Edinburgh, Oxford, IUCAA~(Pune), TIFR~(Mumbai) and UCL~(London) for their kind hospitality and the opportunity to present this work.

\newpage
\appendix
\section{Consistency Relations at Finite Momentum}
\label{sec:AppA}

The soft limits described in \S\ref{sec:softex} were presented in the strict $q \to 0$ limit.  However, since these results follow from current conservation alone, one might expect that they can be extended to finite $q$.  In this appendix, we will demonstrate that this is indeed the case.

\vskip 4pt

\begin{figure}[h!]
   \centering
       \includegraphics[scale =0.5]{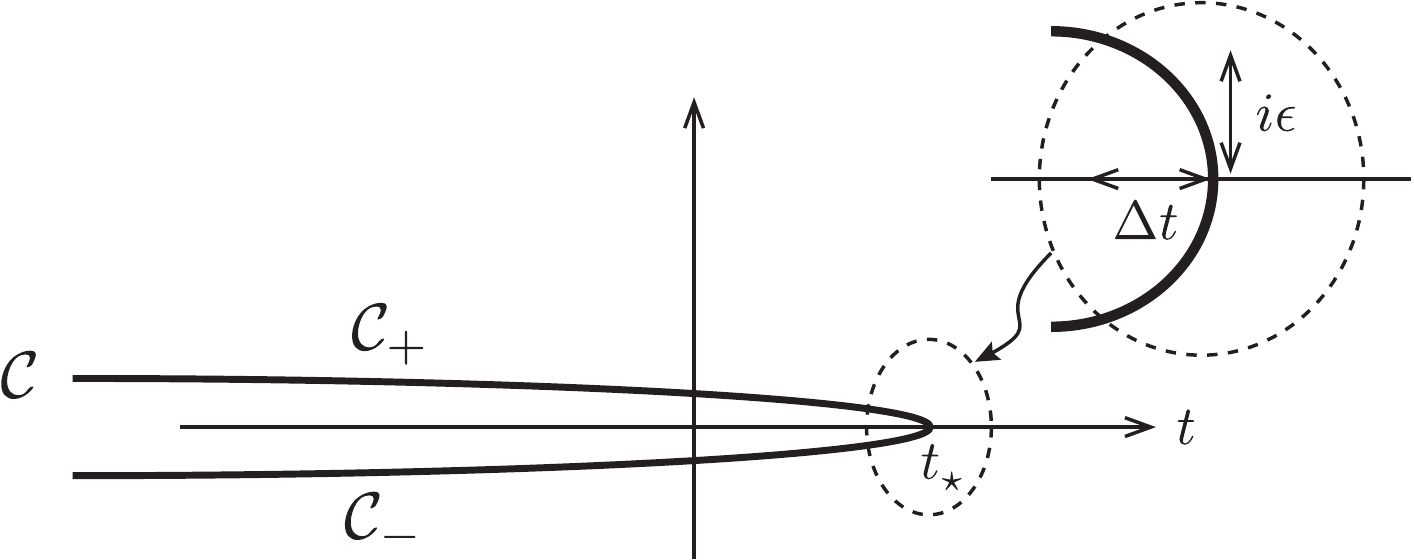}
   \caption{Illustration of the contour employed in the computation of $in$-$in$ correlation functions.}
  \label{fig:contour}
\end{figure}

By definition, any conserved charge is written as an integral of the current over a spatial slice.   Throughout the main text, we have worked directly with the charge $Q \equiv \int \d^3 x\, J^0(\x,t)$ to derive our results.  In particular, we have considered the following commutator
\beq\label{equ:currentdef}
i \langle [Q,\zeta({\y}\hskip 1pt) \zeta( {\z}\hskip 1pt)] \rangle 
\ =\ \langle \delta_{\mathsmaller{Q}} \zeta({\y}\hskip 1pt)\hskip 1pt \zeta({\z}\hskip 1pt) \rangle + \langle  \zeta({\y}\hskip 1pt) \hskip 1pt\delta_{\mathsmaller{Q}} \zeta({\z}\hskip 1pt) \rangle \ , 
\eeq
where it is understood that all fields for which the time dependence isn't shown are evaluated at the same time $t_\star$. 
Eq.~(\ref{equ:currentdef}) follows from current conservation via the Ward identity~\cite{Weinberg:1995mt, Coleman}
\begin{align}
i \, \partial_\mu^{(x)}\langle J^{\mu}(\x,t) \zeta(\y, t_\star) \zeta( \z, t_\star, ) \rangle_{\hskip 1pt T} &\ =\ \delta_{\mathsmaller{\mathcal{C}}}(t-t_\star)\delta({\x}-{\y} \hskip 1pt) \, \langle \delta_{\mathsmaller{Q}} \zeta(\y, t_\star) \hskip 1pt \zeta({\z}, t_\star) \rangle \nonumber\\ &\hspace{1cm} +\, \delta_{\mathsmaller{\mathcal{C}}}(t-t_\star) \delta({\x}-{\z}\hskip 1pt ) \, \langle   \zeta({\y}, t_\star)\hskip 1pt \delta_{\mathsmaller{Q}} \zeta({\z}, t_\star) \rangle  \label{equ:A2}  \ ,
\end{align}
 where the terms proportional to $\delta_{\mathsmaller{\mathcal{C}}}(t-t_\star)$ arise from the time derivative acting on the ``time-ordering" operator $T$.
This feature of the Ward identity deserves a few more words of explanation:
 By definition, the above $in$-$in$ correlation functions are ``time-ordered" on a contour $\mathcal{C}$ that runs from $t = -\infty+i\epsilon$ to $-\infty-i\epsilon$, passing through $t_\star$. Formally, we can write this as
\beq
 \vev{\mathcal{O}_1(t_1)\mathcal{O}_2(t_2)}_{\hskip 1pt T} \, =\, \theta(\lambda_1-\lambda_2) \hskip 1pt \vev{\mathcal{O}_1(t_1)\mathcal{O}_2(t_2)}+\theta(\lambda_2-\lambda_1) \hskip 1pt \vev{\mathcal{O}_2(t_2)\mathcal{O}_1(t_1)}\ ,
 \eeq
 where $\theta$ is the Heaviside function and $\lambda$ is a real number that parametrizes the contour $\mathcal{C}$---e.g. we may choose
 $\lambda =  t-t_\star$ on the ``incoming" contour~${\cal C}_+$ and $\lambda = t_\star-t $ on the ``outgoing" contour~${\cal C}_-$ (see Fig.~\ref{fig:contour}).
  The delta function $\delta_{\mathsmaller{\mathcal{C}}}(t-t_\star)$ on the r.h.s.~of (\ref{equ:A2}) is therefore defined as 
  \beq
  \delta_{\mathsmaller{\mathcal{C}}}(t-t_\star) = \left\{\begin{array}{cc}
  \delta(t-t_\star) & \,\hskip 1pt \text{on $\mathcal{C}_+$}\, ,\\
    -\delta(t_\star-t) & \text{ on $\mathcal{C}_-$}\, .\\
  \end{array}\right.
  \eeq
 If we now integrate
 (\ref{equ:A2}) over the contour $\int^{t_\star -\Delta t -i\epsilon}_{t_\star -\Delta t+i\epsilon} \d t \int \d^3 x \equiv \oint \d t \int \d^3 x$ through $t=t_\star$, we arrive at eq.~(\ref{equ:currentdef}).  We have dropped the integral over $\partial_i J^i$ because it only receives contributions from spatial infinity.

When translated into momentum space, the above treatment is limited to $q = 0$.  However, nothing required that we integrate the Ward identity in this way.  Alternatively, we can multiply eq.~(\ref{equ:A2}) by $ e^{i {\q} \cdot {\x}}$ and then integrate over $\x$ and $t$,
\begin{align}
\oint  \d t \int \d^3 x\  e^{i {\q} \cdot {\x}}\ i\,  \partial_\mu^{(x)} \langle  J^{\mu}( \x,t)  \zeta(\y \hskip 1pt) \zeta(\z \hskip 1pt)  \rangle_{\hskip 1pt T} 
&\ =\ e^{i {\q} \cdot {\y}}\langle \delta_{\mathsmaller{Q}}  \zeta({\y} \hskip 1pt) \hskip 1pt \zeta({\z} \hskip 1pt) \rangle +  e^{i {\q} \cdot {\z}}\langle   \zeta({\y} \hskip 1pt) \hskip 1pt \delta_{\mathsmaller{Q}}  \zeta({\z} \hskip 1pt) \rangle  \ .
\end{align}
Taking the limit $\epsilon, \Delta t \to 0$ and going to momentum space, we find
\beq
i\langle [ J^{0}_{\q}(t_\star) , \zeta_{\k_1} \hskip -1pt \zeta_{\k_2} ] \rangle +  \oint  \d t\ q_i  \langle  J^{i}_{\q}\hskip 1pt(t) \zeta_{\k_1} \hskip -1pt\zeta_{\k_2} \rangle= \langle \delta_{\mathsmaller{Q}}  \zeta_{\q+\k_1} \hskip 1pt \zeta_{\k_2} \rangle+\langle   \zeta_{\k_1} \hskip 1pt \delta_{\mathsmaller{Q}} \zeta_{\q+\k_2} \rangle \ . 
\eeq
Inserting a complete set of states in the first term, one arrives at the identity
\begin{align}
\sum_{n,\p}i \Big[ \langle J^{0}_{\q}(t_\star) | n_\p \rangle \langle n_\p \hskip 1pt | \zeta_{\k_1} \hskip -1pt \zeta_{\k_2}  \rangle - {\rm h.c.} \Big] &\ =\   \langle \delta_{\mathsmaller{Q}} \zeta_{\q+\k_1} \hskip 1pt  \zeta_{\k_2} \rangle + \langle   \zeta_{\k_1} \hskip 1pt \delta_{\mathsmaller{Q}}  \zeta_{\q+\k_2} \rangle   -  \oint \d t\ q_i  \langle  J^{i}_{\q}\hskip 1pt(t) \zeta_{\k_1} \hskip -1pt\zeta_{\k_2} \label{equ:A4} \rangle\ .
\end{align}
Note that this result holds without making any assumptions about the relative or absolute sizes of $q$, $k_1$ and $k_2$.  This isn't surprising, since so far we have only used current conservation which holds at all momenta.  In order to arrive back at eq.~(\ref{equ:currentdef}), we must show that we can indeed neglect the extra term in eq.~(\ref{equ:A4}) in the limit $q \to 0$.  The choice of contour in the complex time domain implies that 
\beq\label{equ:wardq}
\oint \d t \ q_i  \langle  J^{i}_{\q}\hskip 1pt(t) \zeta_{\k_1} \hskip -1pt \zeta_{\k_2} \rangle\ \xrightarrow{\epsilon \to 0} \   \int^{t_\star}_{t_\star -\Delta t } \d t\ q_i \langle [ J^{i}_{\q}\hskip 1pt(t) , \zeta_{\k_1}\hskip -1pt \zeta_{\k_2} ] \rangle \ .
\eeq
Causality requires that 
$ [ J^{i}(\x,t_\star) , \zeta(\y,t_\star) \zeta(\z, t_\star) ]$ vanishes when $\x$ is outside the light-cone\,\footnote{The causal structure of the background will be modified by metric fluctuations.  In all situations of interest in this paper, the fluctuations are small and this effect is negligible \cite{Senatore:2009cf}.  On the other hand, in eternal inflation the presence of large metric fluctuations can modify the causal structure significantly~\cite{Creminelli:2008es}.} 
of $\y$ and $\z$.    For $\x \neq \y,\z$, the operators are space-like separated and the commutator is a proportional to a delta function in position space, or in momentum space $[J^i_{\q} \hskip 2pt , \zeta_{\k} ] = f(q) {\cal O}_{\k+\q}$\hskip 2pt, where $f(q)$ is an analytic function.  Translation invariance ensures that the correlation function $\langle {\cal O}_{\k_1+\q} \hskip 2pt \zeta_{\k_2} \rangle$ can be written in terms of $\k_2$.  Therefore, the final term in (\ref{equ:wardq}) will be analytic in $q$.  Furthermore, invariance under spatial rotations requires that the commutator vanishes at $\q =0$ (using the rotation $\k_1 \to -\k_1$).  These two facts together imply
\beq
\lim_{q \to 0}\, \langle [ J^{i}_{\q}\hskip 1pt(t) , \zeta_{\k_1} \hskip -1pt\zeta_{\k_2} ] \rangle = f(k_1,k_2) \, q^i \ .
\eeq  
Hence, we find 
\begin{align}
\lim_{q \to 0}\ \sum_{n,\p}i \Big[ \langle J^{0}_{\q} | n_\p \rangle \langle n_\p \hskip 1pt | \zeta_{\k_1} \hskip -1pt \zeta_{\k_2}  \rangle - {\rm h.c.} \Big] &\ =\   \langle \delta_{\mathsmaller{Q}} \zeta_{\q+\k_1} \hskip 1pt  \zeta_{\k_2} \rangle + \langle   \zeta_{\k_1} \hskip 1pt \delta_{\mathsmaller{Q}} \zeta_{\q+\k_2} \rangle + {\cal O}(q^2) \ .  \label{equ:A6}
\end{align}
The conclusion that the consistency relation holds at finite $q$ up to corrections suppressed by $q^2$ is extremely important. It ensures that contributions to the l.h.s.~of (\ref{equ:A6}) that vanish like $q^{3- \alpha}$ for $\alpha > 1$ cannot be mimicked by finite $q$ effects in single-field inflation.  See \cite{Creminelli:2011rh} for an alternative derivation of this result.

\newpage
\begingroup\raggedright

\endgroup

\end{document}